\documentclass[%
 reprint,
 amsmath,amssymb,
 aps,
]{revtex4-2}

\usepackage{graphicx}
\usepackage{dcolumn}
\usepackage{bm}
\usepackage{hyperref}

\RequirePackage{snapshot}
\usepackage{amsmath}

\usepackage[capitalise]{cleveref}

\creflabelformat{equation}{#2#1#3}

\providecommand{\tightlist}{%
  \setlength{\itemsep}{0pt}\setlength{\parskip}{0pt}}

\begin{document}

\preprint{APS/123-QED}

\title{Firing statistics in the bistable regime of neurons with homoclinic
spike generation}

\author{Jan-Hendrik Schleimer}
\author{Janina Hesse}
\author{Susana Andrea Contreras}
\author{Susanne Schreiber}

 \affiliation{Institute for Theoretical Biology, Humboldt-Universität zu Berlin}  \affiliation{Bernstein Center for Computational Neuroscience, Berlin}

%
%


\date{\today}

\begin{abstract}
  \textbf{Abstract:} Neuronal voltage dynamics of regularly firing neurons
  typically has one stable attractor: either a fixed point (like in the
  subthreshold regime) or a limit cycle that defines the tonic firing of
  action potentials (in the suprathreshold regime). In two of the three
  spike onset bifurcation sequences that are known to give rise to
  all-or-none type action potentials, however, the resting-state fixpoint
  and limit cycle spiking can coexist in an intermediate regime, resulting
  in bistable dynamics. Here, noise can induce switches between the
  attractors, i.e., between rest and spiking, and thus increase the
  variability of the spike train compared to neurons with only one stable
  attractor. Qualitative features of the resulting spike statistics depend
  on the spike onset bifurcations. This study focuses on the creation of
  the spiking limit cycle via the saddle-homoclinic orbit (HOM)
  bifurcation and derives interspike interval (ISI) densities for a
  conductance-based neuron model in the bistable regime. The ISI densities
  of bistable homoclinic neurons are found to be unimodal yet distinct
  from the inverse Gaussian distribution associated with the
  saddle-node-on-invariant-cycle (SNIC) bifurcation. It is demonstrated
  that for the HOM bifurcation the transition between rest and spiking is
  mainly determined along the downstroke of the action potential -- a
  dynamical feature that is not captured by the commonly used reset neuron
  models. The deduced spike statistics can help to identify HOM dynamics
  in experimental data.
\end{abstract}

\maketitle


\hypertarget{introduction}{%
\section{Introduction}\label{introduction}}

Spike trains recorded from nerve cells vary in their degree of
regularity. Some emit regular tonic pulses like Purkinje cells, other
show very irregular spike trains, such as ``stuttering cells'' or
``irregular spiking cells'' ~{[}1--5{]}. The ubiquitous irregularity of
action-potential (AP) firing in nerve cells has been noted early on
~{[}6{]}, and the functional implications are debated. In some cases,
noise has been deemed an obstacle for reliable responses ~{[}7{]}, while
other studies have conversely highlighted its beneficent involvement in
creating fast or information-optimal responses ~{[}8{]}. Indeed, nervous
systems may well have in store both: situations where irregularity is
facilitating neuronal function ~{[}9{]}, and others where it is
detrimental. While the functional debate is still on, the phenomenology
of irregular spiking has not been completely characterised, let alone
its mechanisms quantitatively understood.

\begin{figure}
\hypertarget{fig:fIcurve_snl}{%
\centering
\includegraphics{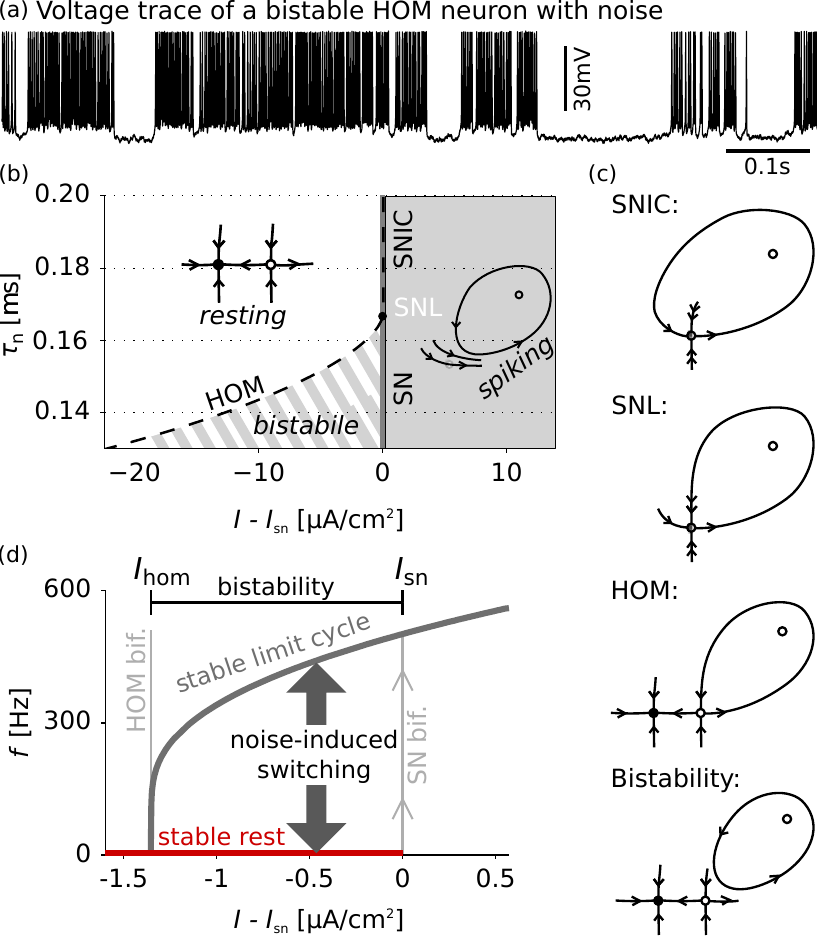}
\caption{Bistability in homoclinic neurons. (a) Voltage trace of a
homoclinic neuron (model definition in Sec. \ref{app:model}, with gating
time constant of \(\tau_n = 0.16\mathrm{ms}\)) driven with
\(I=4.4 \mu\mathrm{A/cm}^2\) and a noise strength of
\(\sigma=24\mathrm{mV}/\sqrt{\mathrm{s}}\) (\(\frac{\sigma^2}2\) is the
voltage diffusion constant in millivolt per second). (b) The homoclinic
regime is reached by decreasing the time scale of the gating variable,
\(\tau_n\). Going from the SNIC to the homoclinic regime, the neuron
passes the codimension-two SNL bifurcation, which is common to all
class-1 excitable neurons. The SNL bifurcation thus acts as a gate to
the bistable, homoclinic spiking regime (shaded area). (c) Phase plots
(gating variable versus voltage) of neurons in different dynamical
regimes, see panel 1(b) for parameter-space organization. (d) Horizontal
transversal of the bifurcation diagram in 1(b) below the SNL point.
Bistability of rest and spiking leads to hysteresis in the
frequency-input curve of homoclinic neurons. Within the bistable region,
noise can switch between rest and spiking.}\label{fig:fIcurve_snl}
}
\end{figure}

The spike patterns emitted by a neuron are influenced by the synaptic
and intrinsic fluctuations in conjunction with the neuron's intrinsic
dynamics. Thus, two major sources of irregularity are conceivable: Some
irregular neurons are simply subject to strong fluctuations, caused by
intrinsic ion-channel noise or by synaptic bombardment, which increase
their interspike interval (ISI) variability ~{[}10{]}. For others, the
deterministic dynamics shows a bistability with coexistence of resting
state and spiking. This leads to increased variability even at moderate
noise levels. For an exemplary voltage trace of the latter case see
\cref{fig:fIcurve_snl}(a). The goal of this article is to characterise
the spiking statistics of such a bistability that arises at a
\emph{saddle-homoclinic-orbit} (HOM) bifurcation, see
\cref{fig:fIcurve_snl}(b) and (c). The HOM bifurcation is a universal
element of the fundamental bifurcation structure of all
conductance-based neuron models ~{[}11,12{]}. HOM excitability shows to
some extent an intermediate behaviour between the classical type I and
type II excitability ~{[}13,14{]}.

The bistability of HOM neurons leads to a hysteresis in the firing-rate
versus input curve, see \cref{fig:fIcurve_snl}(d): Ramping up the input
current, \(I\), the neuron stays at rest until the resting state loses
stability at \({I_{\mathrm{sn}}}\). Conversely, when ramping down the
input current, the neuron remains spiking until the limit cycle (LC)
disappears at \({I_{\mathrm{hom}}}\). The region of bistability extends
from \({I_{\mathrm{hom}}}\) to \({I_{\mathrm{sn}}}\). Consequently,
noise-free deterministic neurons are easily probed experimentally for
hysteresis effects by ramp currents which may then serve as an indicator
for bistable membrane states. In experiments, many biological neuron
types, however, may evade this kind of screening due to their high
degree of stochasticity, in particular when their intrinsic noise causes
them to constantly jump between attraction domains, resulting in a
``mixed state'' that is insensitive to the direction of parameter
change. Yet, the change in interspike interval statistics and the
switching probability between the two attractors that is derived in this
article can still provide insight into the presence of bistability. As
both measures can be estimated from recordings of biological neurons,
they can be used to differentiate neurons in which the irregularity is
solely due to noise from those in which the irregularity is enhanced by
a bistability of the intrinsic dynamics. This might be particularly
interesting when relating single-cell dynamics to the up- and downstates
observed at network level. Previously, up- and downstates on the
single-cell level are modeled by a bistability of two fixed points in
the membrane voltage ~{[}15--18{]}. The here considered setting is
different, with a bistability between a fixed point of the membrane
voltage (the resting state), and the limit cycle (spiking dynamics), yet
if the upstate shows fast spiking behaviour this would be difficult to
distinguish from the two-fixpoints case.

In the space spanned by the three fundamental parameters of
conductance-based neuron models (\emph{i.e.}, membrane leak, capacitance
and input current), a 2D-manifold of HOM bifurcations unfolds from the
degenerate Bogdanov-Takens cusp point, which was proven to generically
occur in these models ~{[}19{]}. Starting with a model showing the
common \emph{saddle-node on invariant cycle} (SNIC) bifurcation at the
creation of the spiking limit cycle, a decrease of the separation of
timescales between voltage and gating kinetics switches the limit cycle
creation to a HOM bifurcation along with the emergence of a bistability
~{[}12{]}, see \cref{fig:fIcurve_snl}(b). The switch in the bifurcation
that creates the spiking limit cycle happens at the codimension-two
\emph{saddle-node loop} (SNL) bifurcation.\footnote{This codim-2
  bifurcation is known as saddle-node separatrix loop ~{[}20{]},
  saddle-node homoclinic orbit ~{[}21{]}, non-central homoclinic loop to
  a saddle-node ~{[}22{]}, orbit flip ~{[}23{]} bifurcation or
  saddle-node loop in some neurobiological context ~{[}24{]}.} It can be
induced by many fundamental parameters in neuronal systems ranging from
leak conductance, capacitance and temperature changes, to modifications
of extracellular potassium concentration ~{[}11{]}. Most importantly for
the present article, between the HOM and the SN branch emerging from the
SNL bifurcation there exists a region of bistable. Besides HOM neurons,
bistability between rest and spiking also occurs in neuron models that
undergo a \emph{subcritical Hopf} bifurcation, followed by a \emph{fold
of limit cycles} at their firing onset. The spike statistics of these
neuron models has previously been explored numerically ~{[}25,26{]}. The
spike statistics for SNIC neurons (upper part of the bifurcation diagram
in \cref{fig:fIcurve_snl}(b)) is well characterised both for the
excitable dynamics, \emph{i.e.}, \(I<{I_{\mathrm{sn}}}\) (fluctuation
driven ~{[}10{]}), and the limit cycle dynamics, where
\(I>{I_{\mathrm{sn}}}\) (mean driven ~{[}27{]}). The statistics in the
bistable region of HOM neurons, however, is less studied and will be
explored in this study. The derivation of the associated interspike
interval statistics fills a gap of knowledge and provides the means to
differentiate alternative underlying bifurcation structures using spike
statistics. In particular, the following analysis focuses on the
situation where the perturbing noise is weak such that the time
evolution is still dominated by the attractors of the nonlinear
dynamical system, with noise only switching between them.

Sec. \ref{sec:mod} introduces the model for which in Sec. \ref{sec:isi}
the interspike interval density is derived. To this end, the stochastic
trajectories are projected onto the unstable manifold of the saddle, see
Sec. \ref{sec:coord}. In this coordinate system both the statistics of
intermittent silence, burst-firing and switching between these regimes
are calculated in Sec. \ref{sec:inter}, \ref{sec:intra} and
\ref{sec:splitprop}, respectively. Estimation of the probability of
switching is discussed and the relation to ISI moments are presented in
Sec \ref{sec:est}. A comparison to a second kind of bistability is drawn
in Sec. \ref{sec:hopf} and the emergence of multimodal ISI densities as
a meens of distinguising between them is addressed, see Sec.
\ref{sec:multimodal}.

\hypertarget{conductance-based-neuron-model-with-homoclinic-bistability}{%
\section{Conductance-based neuron model with homoclinic
bistability}\label{conductance-based-neuron-model-with-homoclinic-bistability}}

\label{sec:mod}

In the bistable regime, transitions between two stable attractors can be
induced by noise fluctuations. The associated transition probability
between the two attractors as well as the resulting spike statistics is
derived in the following for a generic class of conductance-based neuron
models with additive white noise and the limit cycle spike emerging from
a HOM bifurcation. The analysis focuses on HOM neurons that are close to
the SNL bifurcation, which allows for useful assumptions as introduced
later.

The present analysis considers an \(n\)-dimensional conductance-based
neuron model with one voltage dimension, the membrane voltage \(v\), and
a set of \(n-1\) ion channel gates \(a_i\). The dynamics of the state
vector \(\boldsymbol x=[v,a_1,...,a_{n-1}]^\top\in{I\!\!R}^n\) is given
by
\begin{equation}\dot{\boldsymbol x}=\boldsymbol F(\boldsymbol x)+\boldsymbol D(\boldsymbol x)\boldsymbol\xi(t).\label{eq:dynamics}\end{equation}
The additive noise \(\boldsymbol D(\boldsymbol x)\boldsymbol\xi\)
originates from a diffusion approximation of either synaptic or
intrinsic noise sources. The voltage dynamics follows a current-balance
equation
\(F_1(\boldsymbol x)=(I-I_\mathrm{ion}(\boldsymbol x))/{C_{\mathrm{m}}}\),
with membrane capacitance \({C_{\mathrm{m}}}\), and the gates have first
order kinetics, see also Appendix, Sec. \ref{app:model} for model
details. Details on the simulations are also stated in Appendix, Sec.
\ref{app:model}.

\begin{figure}
\hypertarget{fig:separatrix}{%
\centering
\includegraphics{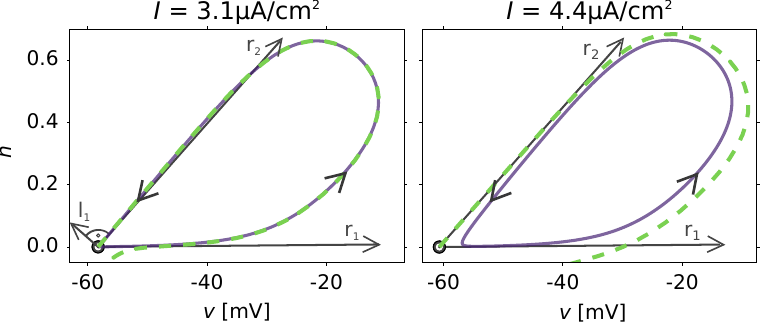}
\caption{Depicted are left and right eigenvector of the saddle as well
as the limit cycle (solid line) and the separatrix (dashed line). (a) At
the HOM bifurcation, the homoclinic orbit overlaps with the separatrix.
(b) For higher input amplitudes, the separatrix is shifted away from the
limit cycle.}\label{fig:separatrix}
}
\end{figure}

The analysis assumes that the model shows a HOM bifurcation from which
the limit cycle spike emerges. A large class of conductance-based neuron
models can be tuned into this regime ~{[}11{]}. In HOM neurons, the
limit cycle (corresponding to tonic firing) arises at
\(I={I_{\mathrm{hom}}}\) from a homoclinic orbit to the saddle, and at
\(I = {I_{\mathrm{sn}}}>{I_{\mathrm{hom}}}\), saddle and stable node
(corresponding to the neuron's resting state) collide in a saddle-node
bifurcation. For inputs in between, with
\({I_{\mathrm{hom}}}< I < {I_{\mathrm{sn}}}\), the stable node and the
limit cycle coexist as two stable attractors, see
\cref{fig:fIcurve_snl}. The state space is divided into the basins of
attraction of the fixpoint and the limit cycle by a separatrix
(\cref{fig:separatrix}).

The analysis furthermore assumes that the noise strength is chosen small
enough such that the spike shape is in first order not affected (the
typical small noise approximation). With this, jumping between spiking
and resting state is only possible close to the separatrix. While the
separatrix is non-local, the following analysis shows that salient
properties of its stochastic transition are given by the linearised
dynamics around saddle and stable node.

The linearised dynamics around fixpoints are given by the Jacobian of
\cref{eq:dynamics},
\(J(\boldsymbol x) = \frac{\partial \boldsymbol F(\boldsymbol x)}{\partial \boldsymbol x}\),
which has \(n\) eigenvalues \(\lambda_{1}, ...,\lambda_{n}\). For
neuronal models undergoing a HOM onset bifurcation, the Jacobian at the
saddle has one simple, positive, real eigenvalue corresponding to the
unstable direction, denoted by \(\lambda_1\in{I\!\!R}\). The other
eigenvalues correspond to stable directions, such that
\begin{equation}\lambda_1>0>\lambda_2\geqslant...\geqslant\lambda_{n}.\label{eq:evals}\end{equation}
The associated orthonormal left and right eigenvectors are denoted by
\(\boldsymbol l_k\) and \(\boldsymbol r_k\), \(k \in [1,...,n]\),
respectively, with \(\boldsymbol l_j\cdot\boldsymbol r_k=\delta_{jk}\),
see \cref{fig:doublewell}(a). Analytical expressions of
\(\boldsymbol l_1\) and \(\boldsymbol r_1\) are given in the Appendix,
in Sec. \ref{app:eigenvectors} for the saddle-node fixpoint, and in Sec.
\ref{app:eigenvectorsSaddle} for the saddle fixpoint.

The statistical properties of the bistable dynamical regime are not yet
sufficiently characterised and will be explored in subsequent sections.

\hypertarget{interspike-interval}{%
\section{Interspike interval}\label{interspike-interval}}

\label{sec:isi}

The following analysis considers the spike train of a HOM neuron in the
bistable region, with \({I_{\mathrm{hom}}}< I < {I_{\mathrm{sn}}}\),
subjected to white noise sufficiently strong to induce jumps between the
two basins of attraction, \emph{e.g.}, \cref{fig:fIcurve_snl}(a).
Between two consecutive spikes, the dynamics can either remain in the
basin of attraction of the limit cycle, or it can visit the basin of
attraction of the fixpoint before eventually returning to the limit
cycle. On average, visiting the fixpoint will induce longer interspike
intervals, because the escape from the resting state requires time in
addition to the duration of the limit cycle.

Because the driving stochastic process is white, the process of
subsequently occurring interspike intervals is renewal, as will be
argued in Sec. \ref{sec:splitprop}. The total interspike interval
density is then a mixture of trajectories that remain on the limit
cycle, and such trajectories with intermittent visits to the fixpoint.
The interspike interval density of these two possibilities are denoted
as

(\emph{i}) the probability \(p_\mathrm{lc}(t)\) that an interspike
interval results from a trajectory staying exclusively on the limit
cycle dynamics, and

(\emph{ii}) the probability \(p_\mathrm{fp}(t)\) that an interspike
interval is composed of some time spent near the resting state in
addition to the time required for the limit cycle spike following the
escape of the fixpoint.

The total interspike interval density is a mixture of both kinds of
trajectories,
\begin{equation}p_\mathrm{isi}(t)=(1-{\varpi})\,p_\mathrm{lc}(t)+{\varpi}\,p_\mathrm{fp}(t),\label{eq:isipdf}\end{equation}
where the factor \(1-{\varpi}\) determines the proportion of intervals
for which the dynamics resides entirely on the limit cycle side of the
separatrix, while \({\varpi}\) is the proportion of intervals that
include time spend on the fixpoint side of the separatrix.

In the following, \({\varpi}\) is called the \emph{mixing factor} or
\emph{splitting probability}. For increasing values of \({\varpi}\),
visits to the fixpoint become more frequent. In spike trains, this is
visible as a larger proportion of long interspike intervals. The
phenomemon of neurons showing strong ISI variability, with a ratio of
long \emph{vs.} short interspike intervals, is sometimes termed
\emph{stochastic bursting}, \emph{stuttering}, \emph{irregularly
spiking}, or \emph{missing spikes} in the experimental literature
~{[}2,4,5{]}.

In the following, the ``ingredients'' to approximate the interspike
interval density in \cref{eq:isipdf} are provided. The mixing factor
\({\varpi}\) is derived in Sec. \ref{sec:splitprop}, and the
probabilities \(p_\mathrm{lc}(t)\) and \(p_\mathrm{fp}(t)\) in Secs.
\ref{sec:intra} and \ref{sec:inter}, respectively. To this aim, the
system is transformed into a coordinate system that facilitates the
analysis (Sec. \ref{sec:coord}).

\hypertarget{projecting-crossings-of-the-separatrix-on-a-double-well-problem}{%
\subsection{Projecting crossings of the separatrix on a double-well
problem}\label{projecting-crossings-of-the-separatrix-on-a-double-well-problem}}

\label{sec:coord}

The observation that most crossings of the separatrix happen along the
downstroke of the action potential (AP) permits in the following to
project the crossings of the separatrix onto a one-dimensional problem.
More specifically, the high-dimensional problem of stochastic
transitions through the \((n-1)\)-dimensional separatrix is reduced to a
one-dimensional double-well escape problem of which the occupancy
statistics are known ~{[}28{]}.

The separatrix between rest and spiking corresponds to the stable
manifold of the saddle fixpoint. At the saddle, the tangent space of the
separatrix is
\begin{equation}\mathcal T=\Big\{\sum\nolimits_{k=2}^{n}\alpha_k\boldsymbol r_k:\alpha_k\in{I\!\!R}\Big\}.\label{eq:tangentspace}\end{equation}
The orthogonal complement is given by the left eigenvector
\(\boldsymbol l_1\in\mathcal T^\perp\), see \cref{fig:doublewell}(a) for
a two-dimensional example.

\begin{figure}
\hypertarget{fig:doublewell}{%
\centering
\includegraphics{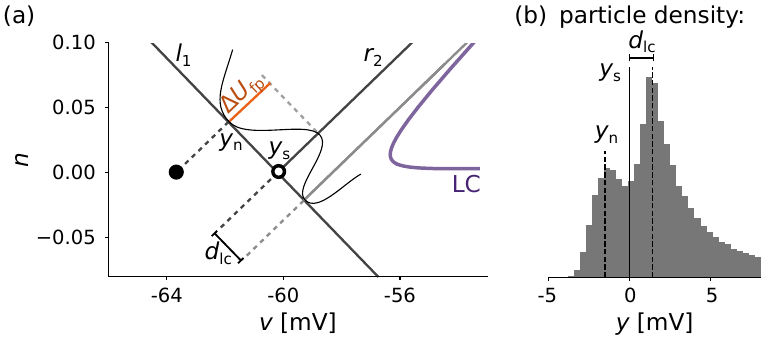}
\caption{Equivalent double-well potential. (a) At the saddle
(\(\circ\)), \(\boldsymbol r_2\) is tangent to the stable manifold. Its
orthogonal complement is \(\boldsymbol l_1\), onto which the node
(\(\bullet\)) and the minimum distance of the limit cycle
(\(d_{\mathrm{lc}}\)) is projected. (b) Simulated particle density in
the projected coordinate \(y\).}\label{fig:doublewell}
}
\end{figure}

For spike onset at \(I={I_{\mathrm{hom}}}\), the separatrix overlaps
with the homoclinic orbit, as both align per definition with the stable
manifold of the saddle. For \(I > {I_{\mathrm{hom}}}\), the limit cycle
detaches from the saddle. The separatrix follows the limit cycle, until
it eventually diverges, see \cref{fig:separatrix}. Along the spike
downstroke, both the limit cycle and the separatrix remain parallel to
the tangent space \(\mathcal T\) for a significant part of the loop, for
details see Appendix, Sec. \ref{app:downstroke}. Most relevant crossings
of the separatrix happen in this region of the state space because

(\emph{i}) due to the slow dynamics in the state space around the
saddle, the limit cycle trajectory spends most of the time close to the
saddle fixpoint, and

(\emph{ii}) the distance between limit cycle and separatrix is minimal
along the spike downstroke, allowing even weak noise deviations to
switch the dynamics between rest and spiking.

In principle, multiple crossings back and forth across the separatrix
are possible, but the final decision is taken when closing in on the
saddle. In the vicinity of the saddle, trajectories on the limit cycle
side of the tangent space \(\mathcal T\) will follow limit cycle
dynamics, while trajectories on the other side of the tangent space
\(\mathcal T\) will visit the stable fixpoint. The decision on which
side of the separatrix a sample path is at a particular time can thus be
read from a projection onto \(\boldsymbol l_1\in\mathcal T^\perp\),
\begin{equation}y(t)=\boldsymbol l_1\cdot(\boldsymbol x(t)-{\boldsymbol x_{\mathrm{s}}}),\label{eq:proj}\end{equation}
where, for simplicity, the dynamics is recentred to the saddle at
\({\boldsymbol x_{\mathrm{s}}}\), such that the saddle is in the
projected coordinates located at \({y_{\mathrm{s}}}=0\). The position of
the stable node is
\({y_{\mathrm{n}}}=\boldsymbol l_1\cdot(\boldsymbol x_\mathrm{n}-{\boldsymbol x_{\mathrm{s}}})\),
see \cref{fig:doublewell}(a). In the following, the convention is used
that \(y>0\) corresponds to the limit cycle side, while \(y<0\) implies
the fixpoint side, corresponding to the rest.

The following analysis uses the minimum distance of the deterministic
limit cycle dynamics to the separatrix,
\begin{equation}d_{\mathrm{lc}}=\mathop{\mathrm{argmin}}_{\boldsymbol x\in\Gamma}\{\boldsymbol l_1\cdot(\boldsymbol x-{\boldsymbol x_{\mathrm{s}}})\},\label{eq:def_dlc}\end{equation}
see \cref{fig:doublewell}(a). Here \(\Gamma\) denotes the invariant set
of the limit cycle. As mentioned above, the minimal distance is
typically reached during the downstroke of the action potential.
\(d_{\mathrm{lc}}\) is the distance in the \(\boldsymbol l_1\)-direction
of the projection along \(\mathcal T\) of the closest point of the limit
cycle to the separatrix.

The projection aims to collapse the decision, whether or not the
fixpoint is visited, into one dimension such that the theory of
double-well potentials can be applied to calculate the occupancy
statistics. A histogram of the projected values, \(y(t)\), from a
simulation shows a bimodal density in \cref{fig:doublewell}(b). Such
bimodal density also appear in the Brownian motion of a particle in a
double-well potential. This motivates the here presented approach to
reduce the properties of stochastic bursting in a high-dimensional
neuron model to a double-well problem:
\begin{equation}\dot y = -U'(y) + \sigma\,\xi(t).\label{eq:noiypotential}\end{equation}
\(y(t)\) here results from the projection of the dynamics onto the
normal direction to the separatrix, as introduced above.

Approximations for the potential \(U(y)\) and the noise strength
\(\sigma\) will be discussed for the different quantities that are
calculated in the following sections.

\hypertarget{splitting-probability}{%
\subsection{Splitting probability}\label{splitting-probability}}

\label{sec:splitprop}

For uncorrelated noise, the series of spike-time events is a renewal
process. After each spike, during the downstroke when the trajectory is
close to the separatrix, the noise in the system operates akin to a
(biased) coin flip that determines if the fixpoint is visited, or if
immediately another round trip on the limit cycle is taken. Hence, the
consecutive decisions from which distribution the spike times are drawn,
\emph{i.e.}, \(p_\mathrm{lc}(t)\) or \(p_\mathrm{fp}(t)\), are Bernoulli
trials (leading to a geometric distribution for the number of spikes per
burst, see \cref{eq:burstlength}). Indeed, the statistics is reduced to
calculating a single parameter: the splitting probability (or mixing
factor) in a double-well potential.

The splitting probability in a double-well potential describes the
probability of a particle that starts at a position in relation to the
barrier to end up in one of the attractors first. In the present case
the particle is initially injected \(d_{\mathrm{lc}}\) away from the
separatrix, see \cref{fig:doublewell}(a). The probabilty of crossing the
barrier and reaching the fixpoint is denoted as \(\varpi\). This
probability can be found by solving the backward Fokker-Planck with the
appropriate boundary conditions ~{[}28{]}. The solution can be expressed
in terms of the steady state density \(p_\mathrm{s}(y)\) as
\begin{equation}\varpi(d_{\mathrm{lc}})=\frac{\int_{d_{\mathrm{lc}}}^{\infty} p_s^{-1}(y)\mathrm{d}y }{\int_{y_\mathrm{fp}}^{\infty} p_s^{-1}(y)\mathrm{d}y}.\label{eq:mixfacdef}\end{equation}

Here, the splitting probability depends on the distance between limit
cycle and separatrix, \(d_{\mathrm{lc}}\), which will be related to the
system parameters in Sec. \ref{sec:LCdistance}.

\begin{figure}
\hypertarget{fig:mixfac_dlc}{%
\centering
\includegraphics{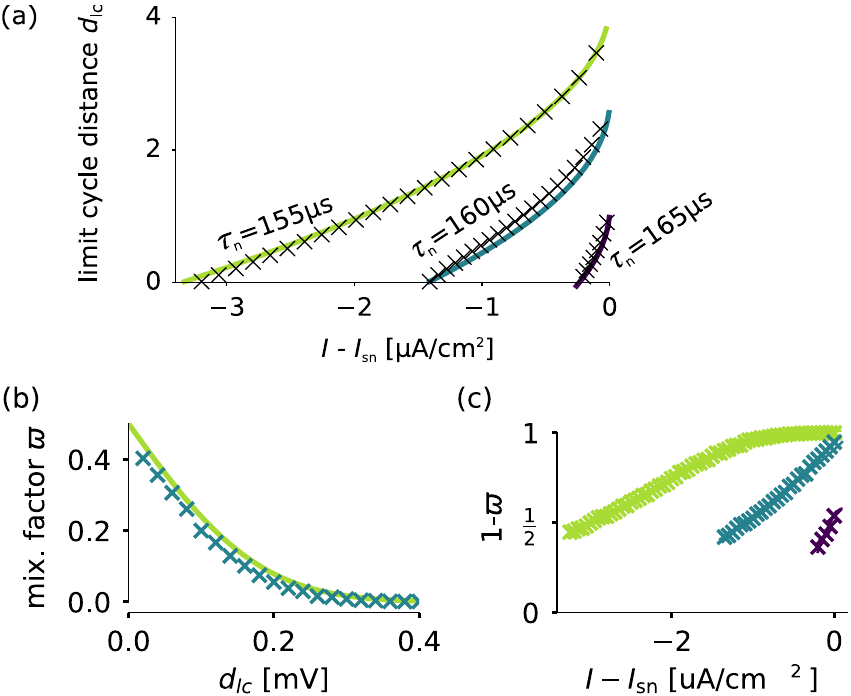}
\caption{Comparison of theoretical prediction (lines) and numerical
simulations (markers) for different gating time constants
\(\tau_\mathrm{n}\)=0.155, 0.16 and 0.165ms. (a) Distance between limit
cycle and separatrix, \(d_{\mathrm{lc}}\), versus input current as given
by \cref{eq:distlc}. (b) Mixing factor \({\varpi}\) as a function of
\(d_{\mathrm{lc}}\). (c) \(1-{\varpi}\) versus input
current.}\label{fig:mixfac_dlc}
}
\end{figure}

The Fokker-Planck equation for the stochastic dynamics of the one
dimensional projected variable \(y(t)\) can always be written in
potential form corresponding to \cref{eq:noiypotential}
\[\partial_t p (y,t) = \partial_y \left[ U'(y) p(y,t) \right] + \frac{\sigma^2}{2}\partial_y^2 p(y,t).\]
The stationary solution \(p_\mathrm{s}\) to this equation can then be
expressed in terms of the potential \(U(y)\) as
\[p_\mathrm{s}(y)= \mathcal{N} \exp\Big(\frac{-U(y)}{2\sigma^2}\Big).\]
Assume that the injection point is not to far from the separatrix which
is at \(y_\mathrm{s}=0\), and that the potential is sufficiently
symmetric around the separatrix. The latter assumption is correct in the
vicinity of the saddle-node bifurcation present in the neuron models
considered here. When \(U(y)\) is smooth, it is possible to assume that
for small \(d_{\mathrm{lc}}\),
\begin{equation}U(y)\approx U(0)+\frac12U''(0)d_{\mathrm{lc}}^2\label{eq:quadratic_simplification}\end{equation}
Assuming \(\sigma\) is small, the limit at \(y=y_{fp}\) tends to
\(-\infty\). With \cref{eq:quadratic_simplification},
\cref{eq:mixfacdef} changes into an expression involving Gaussian
integrals. During the downstroke, the projected limit cycle dynamics
near the separatrix is approximated by \cref{eq:noiypotential}, where
the potential in the direction of \(\boldsymbol l_1\) is
\(U(y)=-\frac{\lambda_1}2y^2\), such that \(U''(0)=-\lambda_1\). The
``mixing noise'' in that dimension is approximated by
\(\sigma^2=\sigma_\mathrm{m}^2=\boldsymbol l_1\cdot{\boldsymbol D}_\mathrm{s} \boldsymbol l_1\),
with the diffusion matrix evaluated at the saddle,
\({\boldsymbol D}_\mathrm{s}={\boldsymbol D}(\boldsymbol x_\mathrm{s})\).
Together this yields
\begin{equation}\varpi=\frac12\bigg(1-\mathop{\mathrm{erf}}\Big(\frac{d_{\mathrm{lc}}\sqrt{\lambda_1}}{2\sigma_\mathrm{m}}\Big)\bigg).\label{eq:splitprob}\end{equation}
Here, \(\mathop{\mathrm{erf}}(\cdot)\) denotes the error function. If
the injection occurs at the separatrix, which corresponds to the
situation when the spiking limit cycle is born from the homoclinic orbit
(\cref{fig:separatrix}), the probability of ending up on either side of
the separatrix is \(1/2\). For increasing distance, the probability of
visiting the fixpoint decays, see inset in \cref{fig:mixfac_dlc}(a),
such that repetitive, burst-like, limit cycle excursions become more
likely.

\hypertarget{limit-cycle-distance-to-the-separatrix}{%
\subsection{Limit cycle distance to the
separatrix}\label{limit-cycle-distance-to-the-separatrix}}

\label{sec:LCdistance}

The limit cycle originates from a homoclinic orbit at
\(I={I_{\mathrm{hom}}}\). As can be seen from the quadratic dynamics in
the centre manifold of the saddle-node, the saddle, and thus the
separatrix, moves as a square-root function of the input current. The
limit cycle position is more invariant, see Appendix, Sec.
\ref{app:inputDependence}. Using \cref{eq:def_dlc}, the distance of the
limit cycle to the saddle in the centre manifold, and thus to the
separatrix, is
\begin{equation}d_{\mathrm{lc}}=\sqrt{\frac{l_{11}}{a\, {C_{\mathrm{m}}}}} (\sqrt{{I_{\mathrm{sn}}}-{I_{\mathrm{hom}}}}
-\sqrt{{I_{\mathrm{sn}}}-I}),\label{eq:distlc}\end{equation} where
\(l_{11}\) is the entry of the left eigenvector \(\boldsymbol l_1\) that
corresponds to the voltage dimension ~{[}29{]}. The factor \(a\) is the
curvature term of the nullclines, and can be determined by ~{[}30,31{]}
\begin{equation}a=\frac12\cdot\boldsymbol l_1\boldsymbol H\boldsymbol r_1\boldsymbol r_1,\label{eq:curvature}\end{equation}
where \(\boldsymbol H\) is the Hessian matrix of the deterministic
dynamics.

\cref{fig:mixfac_dlc}(a) depicts the analytical \(d_{\mathrm{lc}}\) from
\cref{eq:distlc} and the simulated distance of the limit cycle to the
separatrix as a function of the input current. For values of \(I\) away
from the saddle-node, \({I_{\mathrm{hom}}}<I\ll{I_{\mathrm{sn}}}\), the
relation is rather linear. Hence, near the onset of bistability, the
limit cycle distance can be approximated by
\begin{equation}d_{\mathrm{lc}}\approx\sqrt{\frac{l_{11}}{2 a\, {C_{\mathrm{m}}}}} \frac{I-{I_{\mathrm{hom}}}}{\sqrt{{I_{\mathrm{sn}}}-{I_{\mathrm{hom}}}}}.\label{eq:dlcappr}\end{equation}
With these expressions for the distance \(d_{\mathrm{lc}}\), the mixing
factor \({\varpi}\) can be calculated according to \cref{eq:splitprob}.
For comparison, the mixing factor \({\varpi}\) is evaluated in
stochastic simulations. To this end, the relative time spend on the side
of the stable fixpoint and of the limit cycle is detected by recording a
spike when a voltage threshold of -10mV is crossed from below; and
recording a visit to the fixpoint when a two-dimensional threshold is
crossed (crossing the voltage value of the saddle from above and the
value of the \(n\)-variable 5\% above the value corresponding to the
node). The comparison between simulations and the analytical results can
be inspected in \cref{fig:mixfac_dlc}(b).

Next, the probability \(p_\mathrm{lc}(t)\) for staying on the limit
cycle, the probability \(p_\mathrm{fp}(t)\) for visiting the stable
fixpoint, as well as the intra- and interburst statistics are
calculated.

\hypertarget{intraburst-statistics}{%
\subsection{Intraburst statistics}\label{intraburst-statistics}}

\label{sec:intra}

This section determines the probability \(p_\mathrm{lc}(t)\) for staying
on the limit cycle without visiting the fixpoint used in
\cref{eq:isipdf}. From this, the statistics of spikes inside a ``burst''
is derived, \emph{i.e.}, a consecutive sequence of limit cycle
excursions uninterrupted by a crossing of the separatrix into the
attraction domain of the fixpoint.

For trajectories that stay within the basin of attraction of the limit
cycle and a sufficiently small noise amplitude, a phase reduction maps
the process to a one-dimensional Brownian motion in the phase,
\(\theta\), which has constant drift,
\begin{equation}\dot\theta=1/\tau_\mathrm{lc}+\sqrt{2\bar D_\mathrm{lc}}\,\xi(t).\label{eq:phasediffusion}\end{equation}
Here, \(\tau_\mathrm{lc}\) is the intrinsic, deterministic period of the
limit cycle and \(\xi(t)\) a stochastic white-noise process with
effective diffusion matrix \(\bar D_\mathrm{lc}\). The effective
diffusion matrix, \(\bar D_\mathrm{lc}\), is obtained by averaging the
potentially non-stationary noise over the time scale of one interspike
interval with an appropriate weighting function, \(\boldsymbol Z_1\),
that quantifies how susceptible the spike time is to perturbations at a
given phase \(\varphi\) ~{[}32{]}:
\begin{equation}\bar D_\mathrm{lc}=\int_0^1 d\varphi\,\boldsymbol
Z_1(\varphi)\cdot\boldsymbol D(\boldsymbol x_{\mathrm{lc}}(\varphi)) \boldsymbol Z_1(\varphi).\label{eq:phasenoise}\end{equation}
The weighting function is the so-called phase-response curve,
\(\boldsymbol Z_1(\theta)=\boldsymbol\nabla\theta\big|_{\boldsymbol x(\theta)=\boldsymbol x_{\mathrm{lc}}(\theta)}\),
which can be determined numerically or calculated via centre manifold
reductions ~{[}31{]}. Provided that channel or synaptic fluctuations act
on time scales faster than the average limit cycle period, the effective
phase diffusion, \(\bar D_\mathrm{lc}\), quantifies the averaged noise
per interspike interval that causes jitter in the timing of spikes. It
disregards radial excursions due to noise, in particular those that
would cause jumps over the separatrix into the phaseless set (where no
phase is defined). Assuming the intraburst dynamics is governed by the
stochastic phase evolution in \cref{eq:phasediffusion}, the waiting-time
density follows an inverse Gaussian distribution ~{[}27,33{]}
\begin{equation}p_\mathrm{lc}(t)=
\frac{\exp\big(-\frac{(t-\tau_\mathrm{lc})^2}{\tau_\mathrm{lc}^2\,\bar D_\mathrm{lc}\, t}\big)}
{\sqrt{\pi\,\bar D_\mathrm{lc}\, t^3}}.\label{eq:invGauss}\end{equation}
The mean of the distribution, \(\tau_\mathrm{lc}\), is identical to the
deterministic period of the limit cycle. In the case of a homoclinic
neuron, and close to the limit cycle onset (small \(d_{\mathrm{lc}}\))
it scales according to ~{[}29{]}
\begin{equation}\tau_\mathrm{lc}=-\frac1\lambda_1\ln(d_{\mathrm{lc}}).\label{eq:meanintra}\end{equation}
Here, \(d_{\mathrm{lc}}\) is again the distance of the limit cycle to
the separatrix, \emph{cf.} \cref{eq:def_dlc}, which can be expressed in
terms of system parameters in \cref{eq:distlc} and is required to
fulfill \(d_{\mathrm{lc}}\ll1\).

\hypertarget{interburst-statistics}{%
\subsection{Interburst statistics}\label{interburst-statistics}}

\label{sec:inter}

This section develops the probability \(p_\mathrm{fp}(t)\) for
interspike intervals composed of a visit to the resting state fixpoint
and a limit cycle spike used in \cref{eq:isipdf}. The interburst
intervals resulting from fixpoint visits are on average longer than the
intraburst intervals derived in the last section. The corresponding
interspike interval, \(t_\mathrm{fp}\), can be obtained by adding the
time it takes for the trajectory to escape from the fixpoint,
\(t_\mathrm{e}\), and the proceeding time, \(t_\mathrm{lc}\), for a
spike excursion around the limit cycle, to obtain
\(t_\mathrm{fp}=t_\mathrm{e}+t_\mathrm{lc}\). The escape time,
\(t_\mathrm{e}\), from the resting state is described by Poisson
statistics with a Kramer's rate ~{[}10{]}. The required assumption for
Kramer's theory, \emph{i.e.}, that the dynamics be equilibrated around
the resting state, though not perfectly satisfied, appears reasonable
enough, given that the decay time constant of the exponential decay is
correctly described by the escape rate, as previously validated by
comparisons with numerical simulations ~{[}10{]}. However, there is
disagreement in the very short ISIs ~{[}10{]}. Therefore, in the present
case, the escape rate is only supposed to describe the exit over the
separatrix, which is then followed by the time taken for another limit
cycle spike, \(t_\mathrm{lc}\). If the escape and limit cycle dynamics
were to be statistically independent, the waiting time of the complete
interburst statistics \(p_\mathrm{fp}(t)\) would be the convolution of
the escape statistics \(p_\mathrm{e}\) and the additional time
corresponding to the duration of the spike, \(p_\mathrm{lc}\),
\emph{i.e.},
\begin{equation}p_\mathrm{fp}(t)=(p_\mathrm{lc}*p_\mathrm{e})(t)=\int_0^tp_\mathrm{lc}(t-r)p_\mathrm{e}(r)\mathrm{d}r.\label{eq:convPDF}\end{equation}
Note that \cref{eq:convPDF} effectively describes a Poisson neuron with
a refractory period drawn from \(p_\mathrm{lc}\). The assumption of
statistical independence can be motivated by two observations. Firstly,
due to the fast contraction of the stable directions onto the
one-dimensional unstable manifold at the saddle, the trajectories that
leave the stable fixpoint are likely to penetrate the separatrix near
one point. This gives delta-like initial condition for the limit cycle
dynamics and to some extent clears the memory of the preceding
trajectory. Secondly, the noise is uncorrelated.

The interval statistics of the escape, \emph{i.e.}, the Poisson neuron
with Kramer's rate \(1/\tau_\mathrm{e}\), is exponential,
\begin{equation}p_\mathrm{e}(t)=e^{-t/\tau_\mathrm{e}}/\tau_\mathrm{e}.\label{eq:kramer}\end{equation}
The mean interval \(\tau_\mathrm{e}\) is given by the inverse of the
Kramer's rate ~{[}10{]}
\begin{equation}\tau_\mathrm{e}\approx\frac{2\pi}{|\lambda_1|}e^{\Delta{U_{\mathrm{sn}}}/2\sigma^2},\label{eq:escapetime}\end{equation}
where \(\lambda_1\) is the eigenvalue associated with the unstable
manifold of the saddle. \(\Delta{U_{\mathrm{sn}}}\) is the potential
difference between saddle and node,
\(\Delta{U_{\mathrm{sn}}}={U_{\mathrm{sn}}}({y_{\mathrm{s}}})-{U_{\mathrm{sn}}}({y_{\mathrm{n}}})\).
The latter can be approximated in the vicinity of the saddle-node
bifurcation. Saddle and node depart from the saddle-node according to a
square root function, such that locally
\({y_{\mathrm{sn}}}=({y_{\mathrm{s}}}+{y_{\mathrm{n}}})/2\). If
\({y_{\mathrm{s}}}\) and \({y_{\mathrm{n}}}\) have not departed too far
from \({y_{\mathrm{sn}}}\), the potential \(U\) is centrally symmetric
around \({y_{\mathrm{sn}}}\) and hence has no quadratic part
(\emph{i.e.}, the linear dynamics of saddle and node cancel in the
middle). Therefore, the remaining dynamics can be captured in the
following potential:
\[{U_{\mathrm{sn}}}\approx\frac{({I_{\mathrm{sn}}}-I)(y-{y_{\mathrm{sn}}})}{{C_{\mathrm{m}}}}+\frac{a(y-{y_{\mathrm{sn}}})^3}{3},\]
with the factor \(a\) from \cref{eq:curvature}.

The potential difference between saddle and node is hence
\[\Delta{U_{\mathrm{sn}}}\approx(I-{I_{\mathrm{sn}}})({y_{\mathrm{n}}}-{y_{\mathrm{s}}})/{C_{\mathrm{m}}}+ \frac{a}{12}({y_{\mathrm{n}}}-{y_{\mathrm{s}}})^3.\]

\begin{figure}
\hypertarget{fig:fp_eventDuration}{%
\centering
\includegraphics{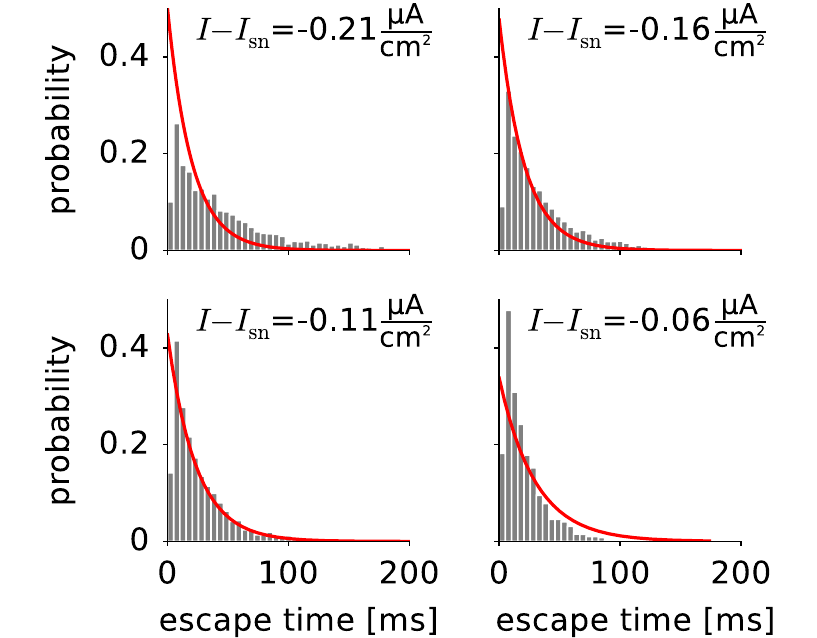}
\caption{Near spike onset, the analytical escape rate (red) fits the
probability density of the escape duration from fixpoint to the
separatrix
(\(\tau_\mathrm{n} = 0.165\mathrm{ms}\)).}\label{fig:fp_eventDuration}
}
\end{figure}

Using this approximation of the potential height in
\cref{eq:escapetime}, the escape time density in \cref{eq:kramer} can be
compared to the simulated neuron. The validity of the approximation can
be inspected in \cref{fig:fp_eventDuration} for different input
currents. With this, all elements of the interspike density in
\cref{eq:isipdf} have been derived.

\begin{figure}
\hypertarget{fig:ISIdensity}{%
\centering
\includegraphics{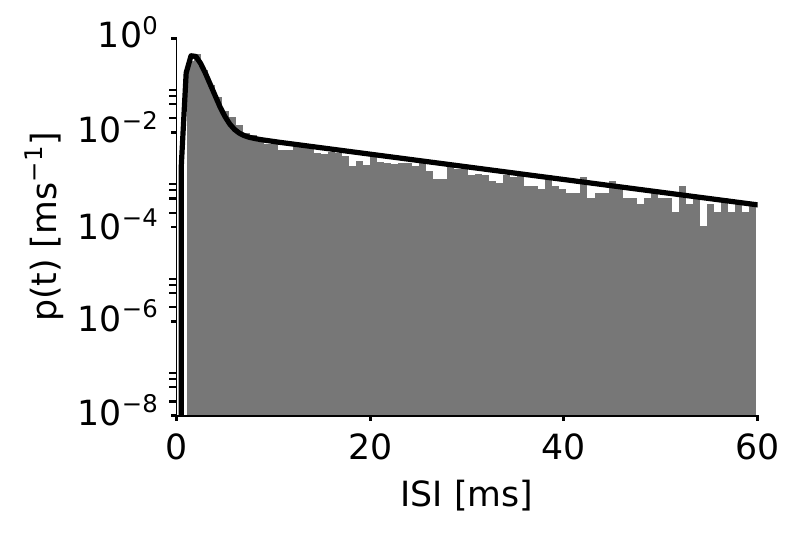}
\caption{Interspike interval density for a numerical simulation with
\(\tau_\mathrm{n} = 0.16\mathrm{ms}\), driven with
\(I =4.4\mu\mathrm{A/cm}^2\) plus current noise with
\(\sigma=0.8\sqrt{\mathrm{ms}}\mu\mathrm{A/cm}^2\). Mean ISI is 4.53 ms,
coefficient of variation is 1.69.}\label{fig:ISIdensity}
}
\end{figure}

The full interspike interval distribution is plotted in
\cref{fig:ISIdensity} for one representative simulation, together with
the analytical prediction.

\hypertarget{burst-length-statistics-and-estimates-of-the-splitting-probability}{%
\subsection{Burst-length statistics and estimates of the splitting
probability}\label{burst-length-statistics-and-estimates-of-the-splitting-probability}}

\label{sec:burstlength}

As argued in Sec. \ref{sec:splitprop}, the sequence of interspike
intervals generated by the present bistable neuron, driven by white
noise, is a renewal process, \emph{i.e.}, after each spike at the
downstroke, the decision from which of the mixture components the
interval is drawn happens irrespective of the previous intervals. Hence,
no serial correlations between intervals are to be expected.
Consequently, the burst length (number of consecutive limit cycle
traverses before crossing the separatrix to the fixpoint) follows a
geometric distribution which only depends on the splitting probability,
\begin{equation}p(k) = {\varpi}(1-{\varpi})^{k-1}.\label{eq:burstlength}\end{equation}

\begin{figure}
\hypertarget{fig:burstlength}{%
\centering
\includegraphics{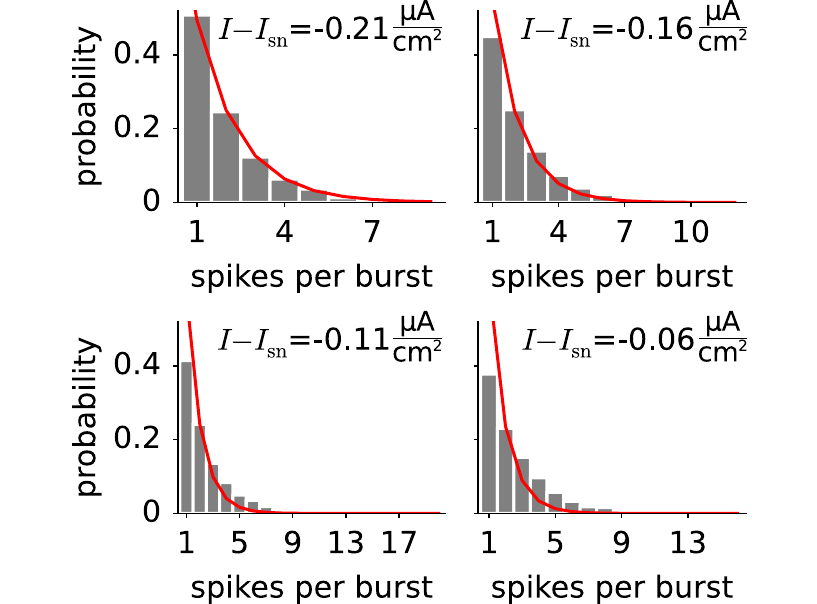}
\caption{Burst-length statistics fitted using a geometric distribution
and the splitting probability from
\cref{eq:splitprob}.}\label{fig:burstlength}
}
\end{figure}

\cref{fig:burstlength} shows a comparison of numerically obtained
burst-length statistics and the theory. This supports the initial
assumption that the distribution of interspike intervals is indeed a
renewal process. The discrepancy between simulations and theory observed
for larger inputs results from the shape of the potential that separates
stable fixpoint and limit cycle. For large inputs, the potential becomes
shallow, such that repeated jumps over the separatrix become more
likely. Even if the dynamics immediately jump back to the limit cycle,
these events are in the simulations counted as fixpoint visits, while
the theory only considers jumps that converge to the fixpoint. This
leads to shorter bursts in the simulation compared to the theoretical
expectations, as observed in \cref{fig:burstlength} in the panel with
\(I - I_{\mathrm{sn}} = -0.06\mu A/cm^2\).

If, for longer experimentally recorded spike trains, histograms of
burst-length distribution are available, the splitting probability
\({\varpi}\) can be inferred as the single parameter that fits \(p(k)\)
to the data.

\hypertarget{moments-and-parameter-estimates}{%
\section{Moments and parameter
estimates}\label{moments-and-parameter-estimates}}

\label{sec:est}

In the presence of noise, hysteresis effects as shown in
\cref{fig:fIcurve_snl}, a distinctive signature of bistability in
deterministic systems, may be obscured. But can bistability still be
detected from stochastic properties of the spike time series? Once
bistability is established, the previous section has identified
multimodality as the distinguishing fact between the bistability
resulting from a saddle-homoclinic orbit bifurcation \emph{versus} a
subcritical Hopf bifurcation.

The splitting probability, \({\varpi}\), may be taken as an indicator
for which regime a neurons is driven,

\begin{enumerate}
\def\labelenumi{(\roman{enumi})}
\item
  \({\varpi}\approx0\to\) mean-driven regime
\item
  \({\varpi}\approx1\to\) fluctuation-driven regime
\item
  \({\varpi}\approx\frac12\to\) bistable neuron
\end{enumerate}

In Sec. \ref{sec:burstlength}, it was surmised that for long enough
spike trains, the mixing factor \({\varpi}\) could be estimated based on
the burst-length statistics in \cref{eq:burstlength}. One may explore
how the moments of the ISI distribution are related to system
parameters. The uncentred moments of the ISI distribution are obtained
from its Laplace transform in \cref{eq:isilaplace} via
\[\nu_k=(-1)^k\frac{\mathrm{d}^k}{\mathrm{d}s^k}P_\mathrm{isi}(s)\Big|_{s=0}.\]
Although the convolution in \cref{eq:convPDF} cannot be evaluated
analytically, its Laplace transform is a simple product of the transform
of the inverse Gaussian distribution of the limit cycle dynamics,
\[P_\mathrm{lc}(s)=\exp\left(\left(1-\sqrt{1+2s\sigma_\mathrm{lc}^2\tau_\mathrm{lc}^2}\right)/\sigma_\mathrm{lc}^2\tau_\mathrm{lc}\right).\]
and that of the exponential distribution, which is
\(P_{\mathrm{fp}}=(1+ s\tau_\mathrm{e})^{-1}\). Together the Laplace
transform of the ISI distribution is
\begin{equation}P_\mathrm{isi}(s)=(1-{\varpi})P_\mathrm{lc}(s)+\frac{{\varpi}P_\mathrm{lc}(s)}{1+s\tau_\mathrm{e}}.\label{eq:isilaplace}\end{equation}

Thus, mean and variance of \(p_\mathrm{isi}(t)\) are given by
\begin{equation}\mu_\mathrm{isi}={\varpi}\tau_\mathrm{e}+ \tau_\mathrm{lc}\label{eq:samplemean}\end{equation}
and
\begin{equation}\sigma_\mathrm{isi}^2=(2-{\varpi}){\varpi}\tau_\mathrm{e}^2+\tau_\mathrm{lc}^3\sigma_\mathrm{lc}^2.\label{eq:samplevar}\end{equation}
For the high firing rates present in HOM neurons with a small
saddle-homoclinic orbit, the mean escape time \(\tau_\mathrm{e}\) is the
longest time scale in the system and can be estimated independently by
fitting a histogram of the largest ISI samples. For low noise,
\(\tau_\mathrm{lc}\) can be estimated as the peak of the ISI histogram.
Then, using \cref{eq:samplemean}, the mixing factor \({\varpi}\) can be
estimated.

\hypertarget{sec:multimodal}{%
\section{Multimodal ISI densities in bistable
neurons}\label{sec:multimodal}}

\label{sec:mulitmodal}

Neuronal bistability at a separatrix connected to the stable manifold of
a saddle is not the only known bistability in single neuron dynamics.
Already in Hodgkin and Huxley's equations for the squid axon a
coexistence of resting and spiking was found for a small parameter range
~{[}34{]}. In that case, for increasing input, a stable and an unstable
limit cycle originate from a fold of limit cycle bifurcation and the
unstable limit cycle subsequently terminates in a subcritical Hopf
bifurcation, which also changes the stability of the fixpoint. ISI
histograms estimated from numerical simulations of the squid model with
noise ~{[}25,26{]}, as well as analytical calculations with simplified
resonate-and-fire type models ~{[}35,36{]}, have suggested the presence
of multimodal peaks in the ISI density. This raises the question if the
kind of bistability in homoclinic neurons treated here can produce
multimodal ISI densities, too, or if this hallmark can be used to
differentiate between the two kinds of bistability?

\hypertarget{hom-case}{%
\subsection{HOM case}\label{hom-case}}

\label{sec:unimodalHOM}

To answer the question of multimodality, the modes of the components of
the mixture are examined. The inverse Gaussian, \(p_\mathrm{lc}(t)\),
has a single mode at
\[\hat t_\mathrm{lc}=\tau_\mathrm{lc}\left( \sqrt{1+\frac{9}{4}\tau_\mathrm{lc}^2\bar D_\mathrm{lc}^2}-\frac{3}{2}\tau_\mathrm{lc}\bar D_\mathrm{lc}\right).\]
The convolution with an exponential kernel does not produce additional
peaks, and hence \(p_\mathrm{fp}(t)\) as defined by the convolution in
\cref{eq:convPDF} is unimodal, too. The derivative of
\(p_\mathrm{fp}(t)\) is
\(\tau_\mathrm{e}p_\mathrm{fp}'(t)=p_\mathrm{lc}(t)-p_\mathrm{fp}(t)\).
If set to zero, it is found that it has a single mode
\(\hat t_\mathrm{fp}\) which satisfies
\begin{equation}p_\mathrm{lc}(\hat t_\mathrm{fp})=p_\mathrm{fp}(\hat t_\mathrm{fp}),\label{eq:modeinter}\end{equation}
\emph{i.e.}, the single mode is located at the crossing of the two
distributions.

The curvature of \(p_\mathrm{fp}\) is given by
\begin{equation}p_\mathrm{fp}''(t)=\frac{1}{\tau_\mathrm{e}}(p_\mathrm{lc}'(t) -
p_\mathrm{fp}'(t)).\label{eq:curvature}\end{equation} The curvature at
the mode is thus given by
\(p_\mathrm{fp}''(\hat t_\mathrm{fp})=p_\mathrm{lc}'(\hat t_\mathrm{fp})/\tau_\mathrm{e}.\)
The curvature is negative because \(\hat t_\mathrm{fp}\) corresponds to
a maximum. Hence, the mode of \(p_\mathrm{fp}(t)\) is to be found on the
declining part of \(p_\mathrm{lc}(t)\), \emph{i.e.},
\(\hat t_\mathrm{lc}<\hat t_\mathrm{fp}\).

The modes of the mixture distribution are confined to lie in the
interval \([\hat t_\mathrm{lc},\hat t_\mathrm{fp}]\). Within this
interval between both individual peaks, \(p_\mathrm{lc}'(t)<0\) and
\(p_\mathrm{fp}'(t)>0\), such that \cref{eq:curvature} implies the
concavity of \(p_\mathrm{fp}(t)\). Let \(\tilde t\) denote the
inflection point of the declining part of the inverse Gaussian
distribution, \(p_\mathrm{lc}(t)\). The distribution
\(p_\mathrm{lc}(t)\) is concave on the interval
\([\hat t_\mathrm{lc},\tilde t]\). Within the interval
\([\hat t_\mathrm{lc},\min(\tilde t,\hat t_\mathrm{fp})]\), both
distributions, \(p_\mathrm{lc}(t)\) and \(p_\mathrm{fp}(t)\), are
concave, which permits no more than a single peak for the mixing
distribution. If the inflection point lies beyond the mode of
\(p_\mathrm{fp}(t)\), \emph{i.e.}, \(\hat t_\mathrm{fp}<\tilde t\), this
implies unimodality of \(p_\mathrm{isi}(t)\). For the other case,
\(\hat t_\mathrm{fp}>\tilde t\), this implies no more than a peak on the
interval \([\hat t_\mathrm{lc},\tilde t]\). For unimodality, it remains
to be shown that the mixing distribution decays on the interval
\([\tilde t, \hat t_\mathrm{fp}]\). Within this interval, let us assume
that \(\tau_\mathrm{e}\) is the longest time scale in the system.
According to \cref{eq:curvature}, the density \(p_\mathrm{fp}(t)\) can
be made arbitrarily flat compared to the derivative
\(p_\mathrm{lc}'(t)\) by increasing \(\tau_\mathrm{e}\). This means that
for sufficiently large \(\tau_\mathrm{e}\), \(p_\mathrm{isi}(t)\) is
within the interval \([\tilde t, \hat t_\mathrm{fp}]\) dominated by the
derivative \(p_\mathrm{lc}'(t)\), and is thus negative with no
possibility for a peak.

Coming back to the question of the modality of bistable homoclinic ISI
density, it can be asserted that for large \(\tau_\mathrm{e}\), which
occur close to \({I_{\mathrm{hom}}}\), and with all other assumptions
used in this article, the ISI density is unimodal. This is in contrast
to at least a large proportion of bistable Hopf neurons and could offer
a way to distinguish these regimes.

\hypertarget{subcritical-hopf-case}{%
\subsection{Subcritical Hopf case}\label{subcritical-hopf-case}}

\label{sec:hopf}

\begin{figure}
\hypertarget{fig:stat_subHopf}{%
\centering
\includegraphics{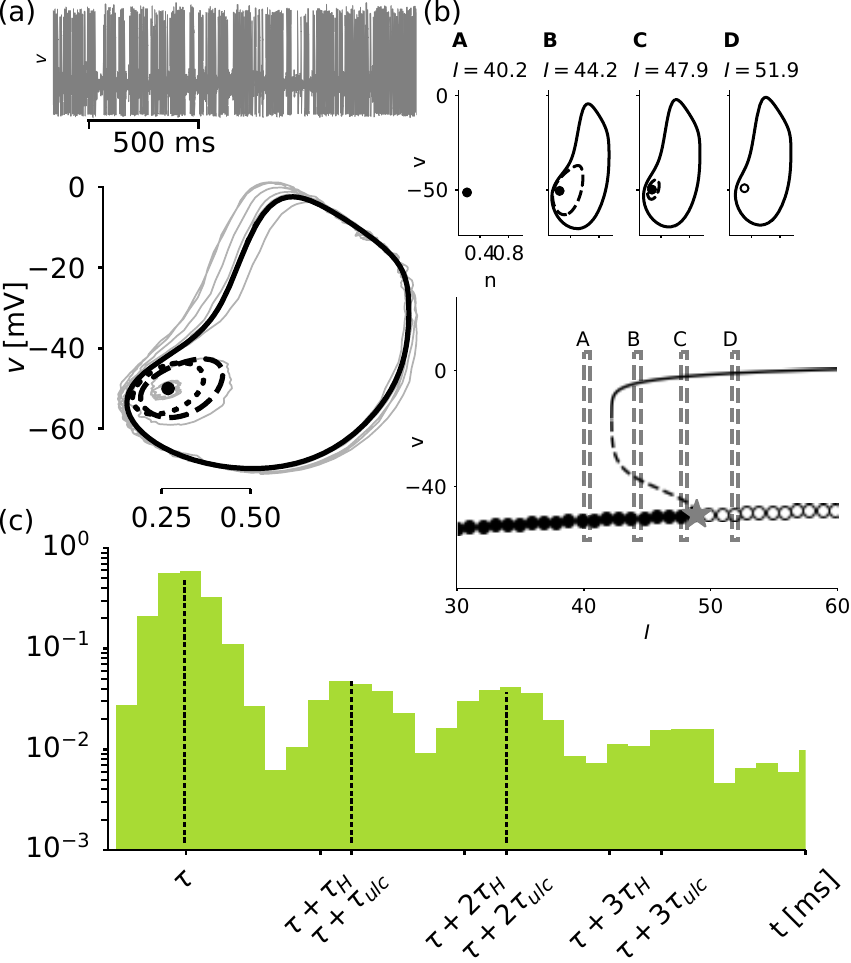}
\caption{(a) Phase portrait and voltage trace of a stochastically
bursting subcritical Hopf neuron are shown in grey. Stable and unstable
limit cycle are shown as solid and dashed black line. Bistability region
is found for DC input parameters between a fold of limit cycles and a
subcritical Hopf bifurcation. The separatrix is an unstable limit cycle.
(b) (A)-(D) show phase portraits before, after and during the emergence
of the unstable limit cycle. Multimodal ISI histogram (c) for noise
parameters derived from the analytically approximated unstable limit
cycle in (a) and shown as the dotted ellipse in (a). Parameter of the
neuron model can be found in ~{[}29{]}, Fig.
6.16.}\label{fig:stat_subHopf}
}
\end{figure}

The second type of bistability in conductance-based neuron models
originates from the case where the stable limit cycle is born together
with an unstable one out of a fold of limit cycles. The unstable limit
cycle shrinks and may either be destroyed by a homoclinic bifurcation to
a saddle ~{[}11{]}, or, as in the Hodgkin-Huxley equations in a
subcritical Hopf bifurcation, which is the case discussed here and shown
in \cref{fig:stat_subHopf}b. The noise-driven dynamics of this latter
case has been investigated in numerical simulations, where the ISI
density was reported as multimodal ~{[}25,26,37{]}. Furthermore, some
geometric considerations about probable exit regions across the unstable
limit cycle have been made ~{[}37{]} and will be discussed in the
following. While at present it is not fully understood how universal
this multimodality is, it can be distinguished from the homoclinic case,
which is not multimodal for sufficiently long escape times from the
fixpoint as argued in Sec. \ref{sec:unimodalHOM}.

In the Hopf case, too, interspike intervals could be categorised into
trajectories cycling around the stable limit cycle and those that visit
the fixpoint region by crossing the circular separatrix given by the
unstable limit cycle as shown in \cref{fig:stat_subHopf}a. Within the
basin of attraction of the stable focus, the dynamics can be linearised,
\begin{equation}\dot{\boldsymbol x}=\boldsymbol J\boldsymbol x+\boldsymbol
B\boldsymbol\xi,\label{eq:linsys}\end{equation} using the Jacobi matrix
at the fixpoint, \(\boldsymbol J\), and given the diffusion matrix,
\(\boldsymbol D=\frac12\boldsymbol B\boldsymbol B^\dagger\). Assuming
there is no focus-to-node transition in the region of bistability
~{[}11{]}, \(\boldsymbol J\) has a pair of conjugate imaginary
eigenvalues with negative real parts. In this case the dynamics shows
noise-induced oscillation (\emph{alias} quasicycles or subthreshold
oscillations) ~{[}38{]}. This class of noisy oscillations that do not
require a deterministic limit cycle can still be described as phase
oscillators using the recently found methods of backward and forward
phases ~{[}39,40{]}. Their average period is given by
\(\tau_\mathrm{H}=2\pi/\omega_\mathrm{H}\) ~{[}40{]}, where
\(\omega_\mathrm{H}\) is the frequency given by the imaginary part of
the Jacobian's eigenvalues at the focus. What determines whether or not
these quasi oscillations are reflected in the ISI density?

For random crossings of the separatrix, one might be inclined to think
that the deterministic definition of phaseless sets ~{[}41{]} implies
that the phases of the spiralling dynamics inside the unstable orbit
will not automatically carry to the phases of the stable limit cycle,
defined by their isocrhons, and thus not influence spike timing
histograms. \emph{A fortiori}, since the isochrons of the stable limit
cycle foliate around the unstable LC ~{[}41,42{]} and thus, in a
deterministic setting, all phases would be available for small
perturbations crossing the separatrix and thus the subthreshold phase
could be scrambled. Yet, the deterministic view is misleading as shown
by numerous simulation studies reporting multimodality ~{[}25,26{]}.
Motivated in part by realistic noise sources such as ion channel
fluctuations, previous considerations have focused on escape from the
fixpoint which is restricted to a region on the unstable LC as a
mechanism for multimodality and found additional peaks ``in all ISI
histograms {[}that{]} have been examined'' ~{[}37{]}. Indeed, the jump
into the unstable limit cycle is typically confined to a region in state
space, because action potentials with a required signal strength (AP
height) need to have a resting state fixpoint that is in one corner of
the AP limit cycle, away from the peak voltage. Due to the closeness of
the AP trajectory and the unstable LC, transitions are likely to occur
in this proximity, in particular if noise were to be restricted to the
voltage dimension. The idea is that each time after completing another
subthrehold cycle there is a probability of jumping out and initialising
a spike at a narrow region in state space. It was previously concluded
that ``the second peak arises when the trajectory {[}\ldots{]} spirals
round in the vicinity of the fixpoint for one cycle, and as the next
cycle starts, it switches back to the stable limit cycle vicinity, thus
creating an ISI roughly twice as long as the period of the stable limit
cycle'' ~{[}37{]}. Actually, the quoted premise would not lead to a peak
at double \(\tau_\mathrm{lc}\) but with that reasoning the modes would
be located at
\[\tau^\mathrm{mode}_\mathrm{H}=\tau_\mathrm{lc}+k\tau_\mathrm{H},\quad\text{for
$k=0,1,..$.}\]

This predicted position of the secondary peaks in the ISI density is,
however, still off as can be seen in \cref{fig:stat_subHopf}c.~A hint on
why, may be inferred from the path simulation in
\cref{fig:stat_subHopf}a showing that as the unstable limit cycle is not
very repellent, trajectories close to it will cycle around it for some
time before switching to the large stable limit cycle of a spike at a
specific region. In that case the mode should be found at multiples of
the period of the unstable limit cycle, \(\tau_\mathrm{s}\),
\[\tau^\mathrm{mode}_\mathrm{s}=\tau_\mathrm{lc}+k\tau_\mathrm{s},\quad\text{for
$k=0,1,..$,}\] which fits better to the data.

In order to highlight that ``regionalised exit'' ~{[}37{]} is not a
prerequisite for multimodality in the ISI, the noise sources in the
system were carefully tuned to render exit over the entire unstable
limit cycle equally likely. Though this might be an unlikely scenario in
a real nerve cell, it is a useful theoretical thought experiment to
understand the requirements for multimodality. The scenario arises for
small unstable limit cycles accompanied by \emph{a fortiori} small
noises with a special structure \(\boldsymbol D\). The diffusion matrix
has to be chosen such that the covariance matrix \(\boldsymbol\Sigma\)
of the stationary solution of \cref{eq:linsys} matches with the geometry
of the surrounding unstable limit cycle. Close to the Hopf bifurcation
the emerging unstable limit cycle can be approximated as an ellipse
~{[}43{]}:
\[\Gamma(t)=\varepsilon(\boldsymbol q_\Re\cos(\omega_\mathrm{H}t)-\boldsymbol q_\Im\sin(\omega_\mathrm{H}t)),\]
where \(\boldsymbol q_\Re\) and \(\boldsymbol q_\Im\) are the real and
imaginary part of the right eigenvector corresponding to the Hopf
bifurcation. The covariance matrix \(\boldsymbol\Sigma\) matches the
geometry of the ellipse if
\(\boldsymbol\Sigma=\boldsymbol q_\Re\cdot \boldsymbol q_\Re^\top+\boldsymbol q_\Im\cdot \boldsymbol q_\Im^\top\).
The diffusion matrix is then chose as ~{[}28,40{]}.
\begin{equation}\boldsymbol D=-(\boldsymbol J\boldsymbol\Sigma+\boldsymbol\Sigma\boldsymbol J^\top).\label{eq:optnoise}\end{equation}
With this choice of \(\boldsymbol D\), exit though each segment of the
unstable limit cycle is equiprobable.

The additional example of uniform exit highlights the fact that
(\emph{i}) the subthreshold backwards phase of the quasi cycles with
associated period \(\tau_\mathrm{H}\), (\emph{ii}) the phase of the
unstable limit cycle with associated period \(\tau_\mathrm{uLC}\) and
(\emph{iii}) the phase associated with the stable limit cycle and
isochrons and the period \(\tau_\mathrm{lc}\) are all connected. This
fact is different to to the saddle case where paths are contracted on
the separatrix into almost a single point and hence the previous
dynamics is not forgotten.

\hypertarget{discussion}{%
\section{Discussion}\label{discussion}}

Interspike-interval distributions are commonly investigated to
characterise spiking behaviour in neurons. Experimentally, these
distributions are easily measured by observing spike trains in response
to step currents or noise injections. Theoretical distributions have
been derived for several types of neuron models, in particular the
Poissonian distribution for fluctuation-driven integrate-and-fire-type
or conductance-based neuron models ~{[}10{]}, and the inverse Gaussian
distribution for mean-driven neurons with a SNIC bifurcation at spike
onset ~{[}27,33,44{]}. Here, the interspike-interval distribution for
neurons with a saddle-homoclinic orbit bifurcation, from which the limit
cycle spike emerges, was derived within the bistable regime. These
neurons show, close to spike onset, a region of bistability between
resting state and spiking, and if the dynamics visits the resting state
between two spikes, particularly long interspike intervals can ensue.

But can the present statistical analysis help to discern HOM/SNL, SNIC
and Hopf bifurcations in recordings? Fitting inverse Gaussian,
exponential or the bistable ISI density, as derived here, to recordings
and comparing the model likelihood can be construed as supportive
evidence for one or the other mechanism. However, for generalised
inverse Gaussian distributions, it was shown that several diffusion
processes can in principle result in the same waiting time distribution,
or, conversely, ISI distributions cannot be uniquely mapped to their
underlying diffusion processes ~{[}45{]}. Therefore, caution is
warranted not to overstress the generality of one's inference about the
mechanistic cause. Nonetheless, features of the ISI density, such as its
skewness, have been related to underlying biophysical processes such as
adaptation currents ~{[}44,46{]}. A question similar in spirit may be
whether neuronal bistability is uniquely tied to the ISI distribution
derived in the present article?

In terms of the underlying bifurcation structure at least one other
scenario giving rise to bistability of spiking and resting has been
described previously and in Sec \ref{sec:hopf}: The subcritical Hopf
bifurcation in association with a fold of limit cycles -- present in the
equations derived for the classical squid axon -- also leads to a region
of bistability and hysteresis ~{[}34{]}. In combination with noise,
numerical investigations ~{[}25,26{]} indicated that the ISI
distribution is multi-modal for the tested parameter combinations. At
present, no parameters have been documented for which multimodality does
not manifest in the ISI density. In the case of simplified
resonate-and-fire models, the ISI distribution has been investigated
analytically and multimodal peaks were confirmed ~{[}35,36{]}. In
contrast, the present manuscript argues for the absence of multimodality
in the homoclinic-type bistability and hence this difference may be
exploited to distinguish both kinds of bistability.

As was argued, bistability in homoclinic neurons can lead to spike time
patterns, which resemble spiking observed experimentally in neurons,
such as ``stuttering cells'' ~{[}1,2{]} or ``irregular spiking cells''
~{[}3--5{]}. Some cells show membrane-voltage bistability in the form of
distinct downstates and upstates ~{[}47{]}. The likelihood of seeing
this dynamics seems to be increased during sleep and certain
anesthetics. The emergence of up-/downstates is associated with altered
concentration dynamics in the intra- and extracellular space. Since the
required time scale separation to induce the SNL bifurcation can also be
achieved by modifying reversal potentials ~{[}48{]}, the resulting
homoclinic bistability may be a putative, alternative mechanism
underlying some of the up- and downstate dynamics observed in neurons.
While up- and downstates have been modeled previously as bistable
fixpoints in an integrate-and-fire like model ~{[}14,18{]}, the
bistability between resting state and spiking dynamics introduced here
is easily implemented in biologically more realistic conductance-based
neuron model.

The emergence of bistability in neurons changes their coding properties,
too. It has been noted that, in the absence of noise, rate coding in
neurons close to a SNIC bifurcation is undermined by undesirable
nonlinearities. More favorable for coding, bursting neurons have been
shown to linearise the rate-tuning curve ~{[}49{]}. Furthermore, in a
network, bistability in the membrane voltage has been shown to increase
the power for certain frequency bands of a population transfer function
~{[}18{]}. In a similar way, the filtering associated with individual
homoclinic neurons can transfer considerably higher frequencies during
the spiking periods ~{[}50{]}. Hence, spike-timing based codes can
benefit from the high-frequency coding arising from the symmetry
breaking that is induced by the switch in spike generation from SNIC to
HOM at the SNL point ~{[}12{]}. The option to visit the fixpoint before
spiking adds to the versatile coding possibilities of these neurons when
explicitly considering their bistability. An open question is if the
interspike-interval distribution of the bistable neuron has favourable
properties similar to the power-law interspike interval density
appearing in some theories of optimal coding ~{[}52{]}: The mutual
information between a fast stimulus and the emitted spike train is
bounded from above by the output entropy of the alphabet (\emph{i.e.},
the spike train entropy) ~{[}53{]}. The spike train entropy is increased
by more diverse spike patterns arising from the stochastic bursting
responses in the bistable regime, compared to the tonic response of a
SNIC neuron. It remains to be shown how the conditional entropy is
influenced, which also contributes to the system's information
transmission rate.

Early theories of spiking network dynamics have, for simplicity, assumed
identical neurons. Networks of such identical, yet highly stochastic,
spiking neurons can generate global rhythms ~{[}54{]}. Even with mild
heterogeneity in the network delays, these phenomena seem to persist
~{[}55{]}. Neuron-intrinsic heterogeneity has also been investigated in
networks of leaky integrate-and-fire (IF) units, using randomly
distributed thresholds. Under a rate coding regime an optimal level of
heterogeneity was suggested ~{[}56{]}. Yet, with leaky IF models lack
the rich bifurcation structure of conductance-based models.
Particularly, heterogeneity in thresholds will not produce the drastic
and critical changes described in the present article. The impact of
heterogeneity in single-neuron parameters that bring about SNL
bifurcations in networks may be surmised to be substantial, but awaits
further studies.

Integrate-and-fire models focus on capturing only the spike upstroke
dynamics while relying on a reset for the spike downstroke. These models
have been used to investigate the influence of rapid upstroke dynamics
(in part influenced by Na\textsuperscript{+} channel dynamics) on coding
~{[}57{]}, and network dynamics ~{[}58{]}. The quadratic IF model can be
derived from the centre manifold reduction of saddle-node bifurcations
~{[}30{]}, and can with an appropriately chosen reset serve as the
``normal form'' of the bifurcation structure in
\cref{fig:fIcurve_snl}(b) ~{[}21,29{]}. Then again, this article shows
the switching dynamics to occur outside the centre manifold dynamics
during the downstroke along the strongly stable direction. Hence, the
window of opportunity for jumping the separatrix is more related to the
timescale of the K\textsuperscript{+} channel dynamics. Nonetheless, the
quadratic IF with a reset above the unstable fixpoint and noise will
produce the same ISI dynamics as derived here and can thus be taken as a
simplified form in network theories of homoclinic bistable neurons.

In summary, the interspike interval distribution derived in this paper
is useful on various levels. It provides an experimental check for
bistability due to homoclinic spike generation, conveys information on
coding properties, and forms the basis for a mean-field theory for
networks with bistable single-neuron dynamics. Translated to other
oscillating systems, the analysis might even inform about homoclinic
bistability beyond the neurosciences.

\hypertarget{acknowledgments}{%
\section{Acknowledgments}\label{acknowledgments}}

Funded by the German Federal Ministry of Education and Research (No.
01GQ1403) and the Deutsche Forschungsgemeinschaft (No.~GRK1589/2).
Authors J.-H.S. and J.H. contributed equally to this work. The authors
would like to thank Paul Pfeiffer for useful comments on the manuscript.

\appendix

\hypertarget{app:model}{%
\section{Model definition}\label{app:model}}

For illustrative purposes, a planar conductance-based neuron model is
considered, but the analysis can be extend to \(n\)-dimensional models
provided the stable manifold of the saddle is \(n-1\) dimensional.

\[{C_{\mathrm{m}}}\dot v = I - I_{\mathrm{ion}}\]

\[\tau_{\mathrm{n}}(v) \dot n = n_\infty(v)-n\]

\(\tau_{\mathrm{n}}(v)\) is a bounded function of the voltage.

For numerical simulations, two-dimensional sodium-potassium neuron
models are used ~{[}29{]}. Model parameters for the neuron with a
saddle-homoclinic orbit bifurcation are given in Table \ref{tab:1} and
for the neuron with a subcritical Hopf bifurcation in Table \ref{tab:2}.
For both models, the ionic current is

\[I_{\mathrm{ion}} = g_\mathrm{L}(E_\mathrm{L}-v) +g_\mathrm{Na}
m_\infty(v) (E_\mathrm{Na}-v)  +
g_\mathrm{K} n (E_\mathrm{K}-v).\]

The activation curves of the gates are

\(m_\infty = \frac{1}{1+\exp(-\frac{v/\mathrm{mV}+25}{5})}\)

\(n_\infty = \frac{1}{1+\exp(-\frac{v/\mathrm{mV}+29}{15})}\)

The gating time constant is independent of \(v\),
\(\tau_{\mathrm{n}}(v)=\tau_{\mathrm{n}}\).

\begin{table}
\caption{Model parameters of the sodium-potassium neuron used for the
simulations of the homoclinic neuron.}
\label{tab:1}
\begin{tabular}{llr}
\hline\noalign{\smallskip}
Parameter &                &
Value  \\
\noalign{\smallskip}\hline\noalign{\smallskip}
Membrane capacitance & $C_\mathrm{m}$    &       $1\mu$F/cm$^2$  \\
Leak reversal potential & $E_\mathrm{L}$    &              $-80$mV  \\
Sodium reversal potential & $E_\mathrm{Na}$   &               $60$mV  \\
Potassium reversal potential & $E_\mathrm{K}$    &              $-90$mV  \\
Maximal leak conductance & $g_\mathrm{L}$    & $8\mathrm{mS/cm}^2$   \\
Maximal sodium cond. &  $g_\mathrm{Na}$   & $20\mathrm{mS/cm}^2$  \\
Maximal potassium cond. & $g_\mathrm{K}$    & $10\mathrm{mS/cm}^2$  \\
Gating time constant & $\tau_{\mathrm{n}}$ & $0.165\mathrm{ms}$ \\
\noalign{\smallskip}\hline
\end{tabular}
\end{table}

\begin{table} \caption{Model parameters of the sodium-potassium neuron used for the simulations of the subcritical Hopf neuron.} 
\label{tab:2} 
\begin{tabular}{llr} 
\hline\noalign{\smallskip} 
Parameter & & Value\\ 
\noalign{\smallskip}\hline\noalign{\smallskip} 
Membrane capacitance & $C_\mathrm{m}$ & $1\mu$F/cm$^2$ \\
Leak reversal potential & $E_\mathrm{L}$ & $-78$mV \\
Sodium reversal potential & $E_\mathrm{Na}$ & $60$mV \\
Potassium reversal potential & $E_\mathrm{K}$ & $-90$mV \\
Maximal leak conductance & $g_\mathrm{L}$ & $1\mathrm{mS/cm}^2$ \\
Maximal sodium cond. & $g_\mathrm{Na}$ & $4\mathrm{mS/cm}^2$ \\
 Maximal potassium cond. & $g_\mathrm{K}$ & $4\mathrm{mS/cm}^2$ \\
 Gating time constant & $\tau_{\mathrm{n}}$ & $1.0\mathrm{ms}$ \\
\noalign{\smallskip}\hline \end{tabular} \end{table}

Voltage dynamics were simulated in the simulation environment brian2
using the internal noise term \emph{xi} ~{[}59{]}.

\hypertarget{app:eigenvectors}{%
\section{Directions of stable and unstable manifold around the
saddle-node}\label{app:eigenvectors}}

For the saddle-node, \(\lambda_1 = 0\), and the associated right
eigenvector is given by

\begin{enumerate}
\def\labelenumi{(\arabic{enumi})}
\tightlist
\item
  \(\boldsymbol r_1=\frac1\kappa\begin{pmatrix}1\\\tau_k\frac{\partial F_k}{\partial v}\\\vdots\end{pmatrix}=\frac1\kappa\begin{pmatrix}1\\\frac{\mathrm{d}}{\mathrm{d}v}{a^{(\infty)}}_{k}(v)|_{v={v_{\mathrm{sn}}}}\\\vdots\end{pmatrix}\),
\end{enumerate}

where \(v={v_{\mathrm{sn}}}\) and \(a_k={a^{(\infty)}}_k(v)\) was used
~{[}31{]}. The eigenvector \(\boldsymbol r_1\) is tangential to the
semi-stable manifold, which corresponds to the centre manifold of the
SNIC bifurcation.

For the saddle-node, the right eigenvectors to
\(\lambda_2, …, \lambda_{n}\) span the tangential space to the stable
manifold. Due to the orthogonality of left and right eigenvectors, the
normal to the tangent space of the stable manifold is given by the left
eigenvector corresponding to the unstable direction, \(\boldsymbol l_1\)
~{[}31{]},

\begin{enumerate}
\def\labelenumi{(\arabic{enumi})}
\setcounter{enumi}{1}
\tightlist
\item
  \(\boldsymbol l_1=\begin{pmatrix}1\\\tau_k\frac{\partial F_0}{\partial a_k}\\\vdots\end{pmatrix}.\)
\end{enumerate}

\hypertarget{app:eigenvectorsSaddle}{%
\section{Directions of stable and unstable manifold around the
saddle}\label{app:eigenvectorsSaddle}}

The analysis of the HOM neuron model assumes that the spike onset lies
in proximity to the SNL bifurcation, such that the saddle at
\({I_{\mathrm{hom}}}\) inherits properties of the saddle-node at
\({I_{\mathrm{sn}}}\). It is shown in the following for a planar model
that this implies similar linearised dynamics along the unstable
manifold of saddle and saddle-node. To this aim, the eigenvectors around
the saddle are expressed as the eigenvectors of the saddle-node, as
given in Sec. \ref{app:eigenvectors}, plus a small term. The closeness
of the saddle to the saddle-node is translated into two mathematical
assumptions: It is assumed that \(\lambda_1\ll1\) at the saddle (because
\(\lambda_1=0\) at the saddle-node), and it is assumed that the voltage
values of saddle and saddle-node are similar,
\(\Delta v = v_\mathrm{saddle} - v_\mathrm{sn}\ll1\).

The linearised dynamics around saddle or saddle-node are given by their
Jacobian. The Jacobian of a two-dimensional system akin to the model in
Sec. \ref{app:model} is given as

\(\boldsymbol J= \begin{pmatrix} \frac{\partial F_0}{\partial v} &  \frac{\partial F_0}{\partial n} \\  \frac{\partial F_1}{\partial v} &  \frac{\partial F_1}{\partial n} \end{pmatrix}  = \begin{pmatrix} a &  b \\  c &  d \end{pmatrix}\).

For a two-dimensional matrix, the eigenvalues are
\(\lambda_{1/2} = 0.5 (a + d \pm E)\) with
\(E = \sqrt{a^2 - 2 a d + 4 b c + d^2}\). The right eigenvector
corresponding to \(\lambda_1\) is \(r_1 = (1, \frac{2 c}{E + a - d})\),
the left eigenvector is \(l_1 = (1, \frac{2 b}{E +a - d})\) (equal to
the right eigenvalue of the transposed matrix).

Expressing \(E\) by \(\lambda_1\) gives
\(r_1 = (1, \frac{c}{-d + \lambda_1})\) and
\(l_1 = (1, \frac{b}{-d + \lambda_1})\).\\

The assumptions allow to approximate the expressions for the
eigenvectors. Because \(d = -1/\tau_n(v)\) is a bounded function of the
voltage, and because of the assumption \(\lambda_1<<1\), one finds
\(\frac{b}{-d + \lambda_1} = \frac{b}{-d} \cdot \frac{1}{1 + \frac{\lambda_1 }{-d}} \approx \frac{b}{-d} \cdot (1 - \frac{\lambda_1 }{-d})\).
Furthermore, the assumptions \(\Delta v <<1\) allows one to develop
\(b(v)\) and \(c(v)\) around \(v_\mathrm{sn}\), with
\(v = v_\mathrm{sn}+\Delta v\).

\[\boldsymbol r_1 \approx (1, \frac{\partial n_\infty}{\partial v}) + (0, \tau_n \lambda_1 ) \]

\[\approx  (1, \frac{\partial n_\infty}{\partial v}(v_\mathrm{sn})) + (0, \frac{\partial^2 n_\infty}{\partial v^2}(v_\mathrm{sn}) \Delta v + \tau_n \lambda_1 ),\]

\[\boldsymbol l_1 \approx (1, \tau_n \frac{\partial F_0}{\partial n}) + (0, \tau_n \lambda_1 ) \]

\[ \approx (1, \tau_n \frac{\partial F_0}{\partial n}(v_\mathrm{sn})) + (0, - g_\mathrm{k} \,p \,n^{p-1} \Delta v + \tau_n \lambda_1 ),\]

where \(b = g_\mathrm{k} \,n^{p} \,(E_\mathrm{k}-v)\) was used for
\(l_1\).

For \(\lambda_1 = 0\), \(\Delta v = 0\), \(\boldsymbol r_1\) and
\(\boldsymbol l_1\) at the saddle-node fixpoint (see Sec.
\ref{app:eigenvectors}) are recovered. This shows that the linear
dynamics at the saddle and at the saddle-node are similar for small
\(\lambda_1\) and \(\Delta v\), in the sense that the linearised
dynamics around the saddle-node are the zeroth order term of an
expansion of the linearised dynamics around the saddle in \(\lambda_1\)
and \(\Delta v\).

\hypertarget{app:inputDependence}{%
\section{Limit cycle downstroke is in first order independent of the
input}\label{app:inputDependence}}

\begin{figure}
\hypertarget{fig:figAppendix}{%
\centering
\includegraphics{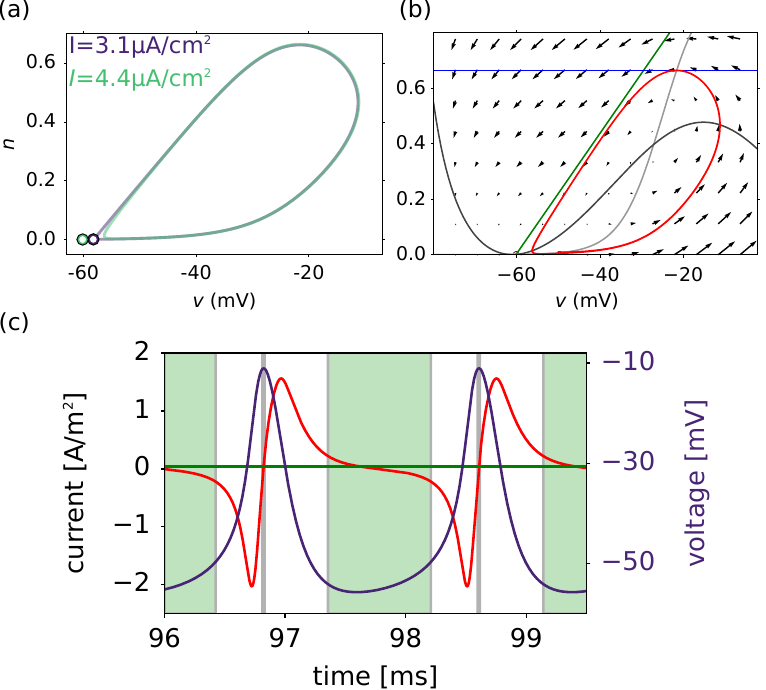}
\caption{Location of the limit cycle. (a) For inputs above spike onset,
the limit cycle position remains relatively stable, while saddle (open
circle), and thus separatrix, move. (b) The downstroke of the limit
cycle aligns with \(\boldsymbol r_2\) (green line). Nullclines are shown
for voltage (dark grey) and gating variable (light grey). (c) During the
spike (violet voltage trace), the ionic current \(I_\mathrm{ion}\) (red)
is orders of magnitude larger than the input current \({I}\) (green
horizontal line). Green background marks the area in which
\(I_\mathrm{ion} < 10 {I}\).}\label{fig:figAppendix}
}
\end{figure}

The simulations in \cref{fig:figAppendix}(a) show that the location of
the limit cycle remains surprisingly constant with an increase in input
current. This can be understood by observing that for the flow on the
limit cycle, the major change in velocity occurs in proximity to the
saddle: The flow on the limit cycle trajectory, \(\Gamma\), is given by
the velocity at each point \((v,n) \in \Gamma\). The speed of the
gating, \(\dot{n}\), is independent of \({I}\). The speed of the
voltage, \(\dot{v} = ({I}+ I_\mathrm{ion})/{C_{\mathrm{m}}}\), depends
on \({I}\), but the influence of \({I}\) is small as long as
\({I}<< I_\mathrm{ion}\). During the spike, ionic currents dominate the
dynamics (\cref{fig:figAppendix}(c)), and \(I_\mathrm{ion}\) has the
same order of magnitude as \({I}\) only in between two spikes,
\emph{i.e.}, close to the saddle (green background in
\cref{fig:figAppendix}(c)). This implies that the only significant
changes in the limit cycle trajectory happen close to the saddle, as
suggested by the simulations of the limit cycle trajectory, see
\cref{fig:figAppendix}(a).

\hypertarget{app:downstroke}{%
\section{At spike onset, the limit cycle downstroke aligns with the
stable eigenvector of the saddle}\label{app:downstroke}}

The derivation in the main text assumes that the downstroke of the limit
cycle follows the stable \(\boldsymbol r_2\) eigenvector of the saddle
at spike onset. This is trivially true for the saddle-homoclinic orbit
in an infinitesimal small environment of the saddle. As shown in the
following, this also holds for a significant part of the dynamics along
the spike downstroke.

The argument relies on the following observations. For the homoclinic
orbit attached to the saddle, two points in the phase plane are known.
(1.) The downstroke of the homoclinic orbit attaches to the saddle. More
precisely, it enters the saddle along its stable manifold tangential to
\(r_2\) (\cref{fig:figAppendix}(b), green line). (2.) At the limit cycle
maximum in the phase plane, \((v(n_\mathrm{max}), n_\mathrm{max})\) with
\(n_\mathrm{max} = \max_t(n_\mathrm{LC}(t))\), the flow is tangential to
the voltage direction, \emph{i.e.}, \(\dot{n} = 0\), such that the
maximum lies on the gating nullcline given by \(n_\infty(v)\)
(\cref{fig:figAppendix}(b)). Because the limit cycle circles around the
unstable node or focus, the limit cycle maximum
\((v(n_\mathrm{max}), n_\mathrm{max})\) lies above the unstable node or
focus. Because voltage and gating nullclines cross at the unstable node
or focus, the limit cycle downstroke, once it has passed
\((v(n_\mathrm{max}), n_\mathrm{max})\), lies above both nullcline. In
consequence, the velocity along the downstroke trajectory points in the
direction of the third quadrant (\emph{i.e.}, to the bottom left), such
that the trajectory approaches \(\boldsymbol r_2\) monotonously. As the
trajectory of the limit cycle downstroke has a horizontal velocity at
\((v(n_\mathrm{max}), n_\mathrm{max})\) and a velocity aligned to
\(r_2\) at the saddle, and approaches \(r_2\) monotonically, the
expectation of smooth trajectories allows to conclude that the
trajectory of the saddle-homoclinic orbit aligns for a significant part
of the limit cycle with \(\boldsymbol r_2\).

\hypertarget{references}{%
\section*{References}\label{references}}
\addcontentsline{toc}{section}{References}

\hypertarget{refs}{}
\leavevmode\hypertarget{ref-gupta_organizing_2000}{}%
{[}1{]} A. Gupta, Y. Wang, and H. Markram, Science \textbf{287}, 273
(2000).

\leavevmode\hypertarget{ref-song_stuttering_2013}{}%
{[}2{]} C. Song, X.-B. Xu, Y. He, Z.-P. Liu, M. Wang, X. Zhang, B.-M.
Li, and B.-X. Pan, PLoS ONE \textbf{8}, (2013).

\leavevmode\hypertarget{ref-galarreta_electrical_2004}{}%
{[}3{]} M. Galarreta, F. Erdélyi, G. Szabó, and S. Hestrin, Journal of
Neuroscience \textbf{24}, 9770 (2004).

\leavevmode\hypertarget{ref-stiefel_origin_2013}{}%
{[}4{]} K. M. Stiefel, B. Englitz, and T. J. Sejnowski, Proceedings of
the National Academy of Sciences \textbf{110}, 7886 (2013).

\leavevmode\hypertarget{ref-mendonca_stochastic_2016}{}%
{[}5{]} P. R. Mendonça, M. Vargas-Caballero, F. Erdélyi, G. Szabó, O.
Paulsen, and H. P. Robinson, eLife \textbf{5}, (2016).

\leavevmode\hypertarget{ref-adrian_basis_1928}{}%
{[}6{]} E. Adrian, \emph{The Basis of Sensation: The Action of the Sense
Organs} (Christophers, 1928).

\leavevmode\hypertarget{ref-schreiber_two_2008}{}%
{[}7{]} S. Schreiber, I. Samengo, and A. V. M. Herz, Journal of
Neurophysiology \textbf{101}, 2239 (2008).

\leavevmode\hypertarget{ref-stemmler_single_1996}{}%
{[}8{]} M. Stemmler, Network: Computation in Neural Systems \textbf{7},
687 (1996).

\leavevmode\hypertarget{ref-ly_noise-enhanced_2017}{}%
{[}9{]} C. Ly and B. Doiron, PLoS ONE \textbf{12}, (2017).

\leavevmode\hypertarget{ref-chow_spontaneous_1996}{}%
{[}10{]} C. C. Chow and J. A. White, Biophysical Journal \textbf{71},
3013 (1996).

\leavevmode\hypertarget{ref-kirst_fundamental_2015}{}%
{[}11{]} C. Kirst, J. Ammer, F. Felmy, A. Herz, and M. Stemmler,
\emph{Fundamental Structure and Modulation of Neuronal Excitability:
Synaptic Control of Coding, Resonance, and Network Synchronization}
(2015).

\leavevmode\hypertarget{ref-hesse_qualitative_2017}{}%
{[}12{]} J. Hesse, J.-H. Schleimer, and S. Schreiber, Physical Review E
\textbf{95}, (2017).

\leavevmode\hypertarget{ref-hodgkin_local_1948}{}%
{[}13{]} A. L. Hodgkin, The Journal of Physiology \textbf{107}, 165
(1948).

\leavevmode\hypertarget{ref-cogno_dynamics_2014}{}%
{[}14{]} S. G. Cogno, S. Schreiber, and I. Samengo, Neural Computation
\textbf{26}, 2798 (2014).

\leavevmode\hypertarget{ref-wilson_origins_1996}{}%
{[}15{]} C. J. Wilson and Y. Kawaguchi, The Journal of Neuroscience: The
Official Journal of the Society for Neuroscience \textbf{16}, 2397
(1996).

\leavevmode\hypertarget{ref-lee_bistability_1998}{}%
{[}16{]} R. H. Lee and C. J. Heckman, Journal of Neurophysiology
\textbf{80}, 583 (1998).

\leavevmode\hypertarget{ref-williams_membrane_2002}{}%
{[}17{]} S. R. Williams, S. R. Christensen, G. J. Stuart, and M.
Häusser, The Journal of Physiology \textbf{539}, 469 (2002).

\leavevmode\hypertarget{ref-wei_impact_2015}{}%
{[}18{]} W. Wei, F. Wolf, and X.-J. Wang, Physical Review E \textbf{92},
(2015).

\leavevmode\hypertarget{ref-kirst_synchronization_2012}{}%
{[}19{]} C. Kirst, Synchronization, Neuronal Excitability, and
Information Flow in Networks of Neuronal Oscillators, PhD thesis,
Niedersächsische Staats-und Universitätsbibliothek Göttingen, 2012.

\leavevmode\hypertarget{ref-schecter_saddle-node_1987}{}%
{[}20{]} S. Schecter, SIAM Journal on Mathematical Analysis \textbf{18},
1142 (1987).

\leavevmode\hypertarget{ref-chow_bifurcation_1990}{}%
{[}21{]} S.-N. Chow and X.-B. Lin, Differential and Integral Equations
\textbf{3}, 435 (1990).

\leavevmode\hypertarget{ref-Aguirre_bifurcations_2015}{}%
{[}22{]} P. Aguirre, SIAM Journal on Applied Dynamical Systems
\textbf{14}, 1600 (2015).

\leavevmode\hypertarget{ref-homburg_homoclinic_2003}{}%
{[}23{]} A. J. Homburg and B. Sandstede, Journal (2003).

\leavevmode\hypertarget{ref-Guckenheimer_bifurcation_1993}{}%
{[}24{]} J. Guckenheimer and J. S. Labouriau, Bulletin of Mathematical
Biology \textbf{55}, 937 (1993).

\leavevmode\hypertarget{ref-tuckwell_analysis_2012}{}%
{[}25{]} H. C. Tuckwell and J. Jost, Physica A: Statistical Mechanics
and Its Applications \textbf{391}, 5311 (2012).

\leavevmode\hypertarget{ref-rowat_isi_2014}{}%
{[}26{]} P. F. Rowat and P. E. Greenwood, Frontiers in Computational
Neuroscience \textbf{8}, (2014).

\leavevmode\hypertarget{ref-Gerstein_random_1964}{}%
{[}27{]} G. L. Gerstein and B. Mandelbrot, Biophysical Journal
\textbf{4}, 41 (1964).

\leavevmode\hypertarget{ref-gardiner_handbook_2004}{}%
{[}28{]} C. W. Gardiner, \emph{Handbook of Stochastic Methods for
Physics, Chemistry, and the Natural Sciences}, 3rd ed (Springer-Verlag,
Berlin ; New York, 2004).

\leavevmode\hypertarget{ref-izhikevich_dynamical_2007}{}%
{[}29{]} E. M. Izhikevich, \emph{Dynamical Systems in Neuroscience} (MIT
Press, 2007).

\leavevmode\hypertarget{ref-ermentrout_parabolic_1986}{}%
{[}30{]} B. Ermentrout and N. Kopell, SIAM Journal on Applied
Mathematics \textbf{46}, 233 (1986).

\leavevmode\hypertarget{ref-schleimer_phase-response_2018}{}%
{[}31{]} J.-H. Schleimer and S. Schreiber, Mathematical Methods in the
Applied Sciences \textbf{41}, 8844 (2018).

\leavevmode\hypertarget{ref-schleimer_spike_2013}{}%
{[}32{]} J. H. Schleimer, Spike Statistics and Coding Properties of
Phase Models, PhD thesis, Humboldt-Universität zu Berlin,
Mathematisch-Naturwissenschaftliche Fakultät I, 2013.

\leavevmode\hypertarget{ref-schroedinger_zur_1915}{}%
{[}33{]} E. Schroedinger, Physikalische Zeitschrift \textbf{16}, 289
(1915).

\leavevmode\hypertarget{ref-rush_potassium_1995}{}%
{[}34{]} M. E. Rush and D. J. Rinzel, Bulletin of Mathematical Biology
\textbf{57}, 899 (1995).

\leavevmode\hypertarget{ref-verechtchaguina_interspike_2007}{}%
{[}35{]} T. Verechtchaguina, I. M. Sokolov, and L. Schimansky-Geier,
Biosystems \textbf{89}, 63 (2007).

\leavevmode\hypertarget{ref-engel_subthreshold_2008}{}%
{[}36{]} T. A. Engel, L. Schimansky-Geier, A. V. M. Herz, S. Schreiber,
and I. Erchova, Journal of Neurophysiology \textbf{100}, 1576 (2008).

\leavevmode\hypertarget{ref-rowat_interspike_2007}{}%
{[}37{]} P. Rowat, Neural Computation \textbf{19}, 1215 (2007).

\leavevmode\hypertarget{ref-brooks_quasicycles_2015}{}%
{[}38{]} H. A. Brooks and P. C. Bressloff, Physical Review E
\textbf{92}, (2015).

\leavevmode\hypertarget{ref-thomas_asymptotic_2014}{}%
{[}39{]} P. J. Thomas and B. Lindner, Physical Review Letters
\textbf{113}, (2014).

\leavevmode\hypertarget{ref-thomas_phase_2019}{}%
{[}40{]} P. J. Thomas and B. Lindner, Physical Review E \textbf{99},
(2019).

\leavevmode\hypertarget{ref-guckenheimer_isochrons_1975}{}%
{[}41{]} J. Guckenheimer, Journal of Mathematical Biology \textbf{1},
259 (1975).

\leavevmode\hypertarget{ref-osinga_continuation-based_2010}{}%
{[}42{]} H. M. Osinga and J. Moehlis, SIAM Journal on Applied Dynamical
Systems \textbf{9}, 1201 (2010).

\leavevmode\hypertarget{ref-kuznetsov_elements_2004}{}%
{[}43{]} Y. A. Kuznetsov, \emph{Elements of Applied Bifurcation Theory}
(Springer Science \& Business Media, 2004).

\leavevmode\hypertarget{ref-schwalger_how_2010}{}%
{[}44{]} T. Schwalger, K. Fisch, J. Benda, and B. Lindner, PLoS
Computational Biology \textbf{6}, (2010).

\leavevmode\hypertarget{ref-barndorff-nielsen_first_1978}{}%
{[}45{]} O. Barndorff-Nielsen, P. Blesild, and C. Halgreen, Stochastic
Processes and Their Applications \textbf{7}, 49 (1978).

\leavevmode\hypertarget{ref-fisch_channel_2012}{}%
{[}46{]} K. Fisch, T. Schwalger, B. Lindner, A. V. M. Herz, and J.
Benda, Journal of Neuroscience \textbf{32}, 17332 (2012).

\leavevmode\hypertarget{ref-loewenstein_bistability_2005}{}%
{[}47{]} Y. Loewenstein, S. Mahon, P. Chadderton, K. Kitamura, H.
Sompolinsky, Y. Yarom, and M. Häusser, Nature Neuroscience \textbf{8},
202 (2005).

\leavevmode\hypertarget{ref-hahn_bistability_2001}{}%
{[}48{]} P. J. Hahn and D. M. Durand, Journal of Computational
Neuroscience \textbf{11}, 5 (2001).

\leavevmode\hypertarget{ref-macgregor_phasic_2012}{}%
{[}49{]} D. J. MacGregor and G. Leng, PLoS Computational Biology
\textbf{8}, (2012).

\leavevmode\hypertarget{ref-brown_phase_2004}{}%
{[}50{]} E. Brown, J. Moehlis, and P. Holmes, Neural Computation
\textbf{16}, 673 (2004).

\leavevmode\hypertarget{ref-schleimer_coding_2009}{}%
{[}51{]} J.-H. Schleimer and M. Stemmler, Physical Review Letters
\textbf{103}, 248105 (2009).

\leavevmode\hypertarget{ref-tsubo_power-law_2012}{}%
{[}52{]} Y. Tsubo, Y. Isomura, and T. Fukai, PLoS Computational Biology
\textbf{8}, (2012).

\leavevmode\hypertarget{ref-strong_entropy_1998}{}%
{[}53{]} S. P. Strong, R. Koberle, R. R. de Ruyter van Steveninck, and
W. Bialek, Physical Review Letters \textbf{80}, 197 (1998).

\leavevmode\hypertarget{ref-brunel_fast_1999}{}%
{[}54{]} N. Brunel and V. Hakim, Neural Computation \textbf{11}, 1621
(1999).

\leavevmode\hypertarget{ref-brunel_dynamics_2000}{}%
{[}55{]} N. Brunel, Journal of Computational Neuroscience \textbf{8},
183 (2000).

\leavevmode\hypertarget{ref-mejias_optimal_2012}{}%
{[}56{]} J. F. Mejias and A. Longtin, Physical Review Letters
\textbf{108}, (2012).

\leavevmode\hypertarget{ref-fourcaud-trocme_how_2003}{}%
{[}57{]} N. Fourcaud-Trocmé, D. Hansel, C. van Vreeswijk, and N. Brunel,
The Journal of Neuroscience \textbf{23}, 11628 (2003).

\leavevmode\hypertarget{ref-monteforte_dynamical_2010}{}%
{[}58{]} M. Monteforte and F. Wolf, Physical Review Letters
\textbf{105}, (2010).

\leavevmode\hypertarget{ref-stimberg_equation-oriented_2014}{}%
{[}59{]} M. Stimberg, D. F. M. Goodman, V. Benichoux, and R. Brette,
Frontiers in Neuroinformatics \textbf{8}, (2014).


\end{document}